\documentclass[12pt]{article}
\usepackage{amssymb,amsmath,epsfig}
\usepackage{array,multirow,graphicx}
\allowdisplaybreaks

\begin{document}

\title{\bf Stable Bounded Excursion Gravastars with Regular Black Holes}
\author{M. Sharif \thanks {msharif.math@pu.edu.pk} and Faisal Javed
\thanks{faisaljaved.math@gmail.com}\\
Department of Mathematics, University of the Punjab,\\
Quaid-e-Azam Campus, Lahore-54590, Pakistan.}

\date{}
\maketitle

\begin{abstract}
This paper explores the possible existence of stable regions of
bounded excursion gravastars in the background of regular black
holes (Bardeen and Bardeen-de Sitter black holes). For this purpose,
we match internal de Sitter geometry with an external regular black
hole through an intermediate shell with an equation of state. The
matter surface located at the shell greatly affects the dynamical
configuration of the shell. We consider stiff and dust fluids to
discuss the outcomes of the developed structures. We then determine
the critical values of physical parameters at which the shell
neither collapses nor expands for both cases with different choices
of external and internal geometries. It is found that there does not
exist any possible region of stable bounded excursion gravastar for
the interior flat and exterior regular geometries. However, the
stable bounded excursion gravastar is obtained for the interior de
Sitter and exterior regular black holes with a suitable choice of
physical parameters for both stiff and dust shells. We conclude that
the possibility of the existence of both bounded excursion
gravastars as well as black holes cannot be excluded in the
dynamical model.
\end{abstract}

\section{Introduction}

The gravitational collapse has been one of the interesting research
fields which lead to the formation of different compact objects such
as white dwarfs, neutron stars, naked singularities, and black holes
(BHs). In the light of theoretical and observational advances, there
is often a set of paradoxical issues surrounding the BHs that
inspires scholars to try other solutions where massive stars are
without horizons at their endpoints of gravitational collapse (Wald
2001). Gravastars (Mazur and Mottola 2001; 2004), Bose superfluid
(Chapline et al 2003) and black stars are a few examples of such
physical models. Within such simulations, gravastars have obtained a
particular interest in recent times, in part, owing to the strong
relationship between the cosmological constant and the expanding
universe (Copeland et al 2006), even though the presence of these
stars may be limited very narrowly on observational grounds. A
gravastar (gravitation expectation star) is an astronomic object
proposed as an alternative to BH with no event horizon and
singularity.

Mazur and Mottola (2004) considered the Visser cut and paste
approach to develop a gravastar model from the joining of exterior
Schwarzschild BH with interior de Sitter (dS) spacetime. This model
can be described in three different regions with the specific
equation of state (EoS). The first region is denoted as an interior
($0\leq r<r_1$), second is the intermediate ($r_1< r<r_2$) and third
is referred to as an exterior region ($r_2<r$). In the first zone, a
repulsive force is generated in the intermediate region by the
isotropic pressure ($p=-\sigma$, where $\sigma$ is the energy
density). The intermediate region is assumed to be shielded by
ultra-relativistic plasma and fluid pressure ($p=\sigma$). The
exterior region is supported by the vacuum solution of the field
equations with zero pressure ($p=0=\sigma$). This provides an
effective thermodynamic solution with small fluctuations and maximum
entropy.

The thin-shell matter surface produces an adequate amount of
pressure to counteract the strength of gravity effects that maintain
its configuration stable. Visser and Wiltshire (2004) found the
simplest model to represent the Mazur-Mottola scenario by combining
external and internal geometries using the cut and paste method.
They evaluated the stable structure of gravastar with a suitable
choice of EoS for the transition layers. There are two different
approaches to studying the dynamics and stability of thin-shell. The
first is to prescribe a potential $V(a)$ and leave the EoS of the
shell as derived. The second is to prescribe an EoS of the shell and
leave the potential $V(a)$ as derived. They followed the first
approach and studied in detail the case where $V(a_0)=0$,
$V'(a_0)=0$ and $ V''(a_0)>0$. Here, prime denotes the ordinary
derivative to the indicated argument. The developed structure is
said to be stable if there exists shell radius $a_0$ for which the
above conditions are verified. There is a less stringent notion of
stability known as bounded excursion models, in which there exist
two radii $a_1$ and $a_2$ with ($a_1<a_2$) such that $V(a_1)=0$,
$V'(a_1)<0$ and $V(a_2)=0$, $V'(a_2)>0$. Carter (2005) extended this
concept by the joining of interior dS spacetime and exterior
Reissner-Nordstr\" om (RN) BH. Horvat et al. (2009) introduced a
theoretical electromagnetic gravastar model and investigated the
effects of charge on stable gravastar configuration. The physical
features such as entropy, proper length, and energy content of
charged gravastar in (2+1)-dimensional spacetime were analyzed by
Rahaman et al. (2012a; 2012b).

Timelike thin-shells in spherically symmetric static spacetimes are
the most interesting cosmological objects which can be constructed
in general relativity. Such models of cosmological objects have been
used to analyze some astrophysical phenomena such as gravitational
collapse and supernova explosions. The seminal work of Israel (1966;
1967) provided a concrete formalism for constructing timelike
shells, in general, by gluing two different manifolds at the
location of the thin-shell. Self-gravitating thin-shells are the
solutions of a given gravitational theory describing two regions
separated by an infinitesimally thin region where the matter is
confined. It is important to determine whether relevant thin-shell
configurations are stable, both thermodynamically and dynamically.

Many researchers have discussed the dynamics and stable
configuration of thin-shell and thin-shell wormholes (WHs) by using
radial perturbation with different matter distributions. Brady et
al. (1991) investigated linear stability of thin-shell that connects
inner flat and outer Schwarzschild BH through radial perturbation.
Martinez (1996) studied thermodynamical stability of such a
geometrical structure. Later, Mazharimousavi et al. (2017) explored
stable configuration of such systems by using variable equation of
state (EoS). LeMaitre and Poisson (2019) observed stability of this
geometrical structure in Newtonian and relativistic gravity. They
found a link between the existence of a maximum mass along a
sequence of equilibrium configurations and the onset of dynamical
instability. Recently, Bergliaffa et al. (2006) examined their
linear and thermodynamical stability by considering barotropic and
also for EoS of the type $p=p(\rho(a))$.

A requirement for a traversable WH is that the throat needs to be
held open and lined with a stress-energy density that violates the
energy conditions of general relativity. The source of such
stress-energy tensor is often called exotic matter. To simplify the
calculations and investigations of the stability of traversable WHs,
Visser (1989) introduced a cut and paste method where the throat is
cut out and the two mouths are pasted together. In this procedure,
the exotic matter reduces in an infinitesimally thin-shell which
connects two equivalent copies of BH spacetimes. For instance, the
components of stress-energy tensor of thin-shell WHs allows us to
conjure up different EoS and also study how these affect the WH
stability. There is a large body of literature that explore the
stable characteristics of thin-shell WHs in the background various
singular and non-singular BH geometries (Halilsoya et al 2014; Lobo
et al 2004; Eiroa and Simeone 2007; Mazharimousavi et al 2010; Dias
and Lemos 2010; Amirabi et al 2013).

Rocha and his collaborators (2008a; 2008b) presented the prototype
gravastar model by using external Schwarzschild BH and internal dS
spacetime. They found that models sometimes represent stable bounded
excursion gravastars, in which thin-shell oscillates between two
final radii, otherwise, they collapse until BHs are formed. Hence,
the possibility of the existence of a gravastar model cannot be
excluded in such dynamical models. Chan et al. (2009a) proposed a
dynamical model of prototype gravastars filled with phantom energy.
It is found that the developed structure can be a BH, stable,
unstable, or bounded excursion gravastar for various matter
distributions at thin-shell. Later, this work was extended for the
choice of exterior Schwarzschild-dS and RN spacetimes with interior
dS region with specific EoS. They found that stable bounded
excursion gravastar is obtained in the presence of exterior
cosmological constant and charge (Chan et al 2009b; 2010).

Lobo and Arellano (2007) studied the gravstar models with non-linear
electrodynamics. G$\acute{a}$sp$\acute{a}$r and R$\acute{a}$cz
(2010) observed the stability of gravastars through the inelastic
collision of their surface layer with a dust shell. Horvat et al.
(2011) considered the gravastar with continuous pressure and
examined stability through the conventional Chandrasekhar approach.
Lobo and Garattini (2013) studied the stability of noncommutative
thin-shell gravastar and found that stable regions must be located
near the predicted location of the event horizon. \"{O}vg\"{u}n et
al. (2017) developed the gravastar model from the matching of
external charged noncommutative BH with internal dS manifold. They
noticed that the established framework satisfies the null energy
condition and displays stable behavior for certain acceptable values
of the physical parameter close to the predicted horizon.

Banerjee et al. (2018) developed thin-shell gravastar solutions with
interior dS and exterior RN BH for a certain range of parameters
where the metric potentials and electromagnetic fields are related
in some particular relation called Guilfoyles solution. They also
observed energy conditions and linearized stability of the developed
structure. In the backgrounds of Bardeen/Bardeen-dS BHs, we have
examined the stable regions of thin-shell gravastars through radial
perturbation (Sharif and Javed 2020). It is found that stable
regions decrease for large values of charge and increase for higher
values of the cosmological constant. There is a large body of
literature (Lobo 2006; DeBenedictis et al 2006; Horvat and
Iliji$\acute{c}$ 2007; Chan et al 2011a; Horvat et al 2011; Chan et
al 2011b; Banerjee and Rahaman 2016; Ghosh and Rahaman 2017; Shamir
and Ahmad 2018; Yousaf et al 2019; Sharif and Waseem 2019; Bhattiet
al 2020; Das 2020; Yousaf 2020) that explores the configuration
thin-shell gravastars in the context of different scenarios.

The establishment of general relativity by Einstein was a great
success of physics in the last century. However, there still remain
a few difficulties out of which the existence of singularity is an
intrinsic problem of general relativity. Regular black hole (BH) is
a concept produced out of multiple attempts to establish an
understandable interior structure for BH by avoiding singularity
inside it. The study of BHs without singularities can help us to
understand the role played by singularities in astrophysics. The
study on global regularity of BHs has attained remarkable importance
to understand the final state of initially regular configurations.
None of the regular BHs are exact solutions to the Einstein field
equations without any physically reasonable source.

The above-mentioned stable/unstable bounded excursion gravastar
models are constructed in the background of exterior geometries with
singularities. For the regular BHs (BHs without singularity), such
geometrical structures have not been explored so far. Regular BHs
are more interesting compact objects due to their regular center.
Bardeen obtained a BH solution without singularity well-known as
Bardeen BH (Bardeen 1968). Bardeen regular BHs are charged
spacetimes which behave as ordinary RN BH solutions and the
existence of these solutions does not contradict with the
singularity theorems. Afterward, some more models of regular BHs
were proposed (Ay$\acute{o}$n-Beato and Garc$\acute{i}$a (ABG) 1998;
Bronnikov 2000; Hayward (2006)). Ay$\acute{o}$n-Beato and
Garc$\acute{i}$a (2000) also proved that the Bardeen BH can be
interpreted as a gravitationally collapsed magnetic monopole arising
in a specific form of nonlinear electrodynamics. Hence, the
stress-energy tensor of the nonlinear electrodynamics is considered
as the source for the Einsteins field equations. Bardeen BH can be
used in dS and AdS backgrounds by Fernando (2017).

Regular BHs and the presence of cosmological constant motivate us to
observe the geometrical structure and stable configuration of
gravastars. Therefore, we consider Bardeen and Bardeen-dS BHs to
develop the geometrical structure of gravastars. This paper is
devoted to present the formalism of a gravastar shell in the
background of regular BHs through the cut and paste technique. We
explore the existence of stable bounded excursion gravastar for
suitable values of physical parameters. The paper has the following
format. Section \textbf{2} provides the construction of a gravastar
shell and also evaluates the respective potential function by using
the equation of motion with suitable EoS. In section \textbf{3}, we
investigate the developed structure for stiff and dust fluids in the
presence of different exterior and interior geometries with critical
values of physical parameters. Finally, we summarize our results in
the last section.

\section{Gravastar Model with Regular Black Holes}

The line elements of lower and upper manifolds of gravastar shell in
(3+1)-dimensions with coordinates $(t,r,\theta,\phi)$ can be
expressed as
\begin{equation}\label{1}
ds^2=g_{\mu\nu}dx^\mu
dx^\nu=-k(1-\lambda_n(r))dt^{2}+(1-\lambda_n(r))^{-1}dr^{2}+r^2d\Omega^2,
\end{equation}
where $n=i,e$ denote the interior ($i$) and exterior ($e$) regions
with the time coordinate scaling constant $k>0$ and metric of unit
2-sphere is $d\Omega^2=d\theta^2+\sin^2\theta d\phi^2$. Here,
$\lambda_n(r)$ represents the compactness function that can be
written as
\begin{equation}\label{2}
\lambda_n(r)=2m_n(r)/r=2/r\int_{0}^{r}4\pi
\bar{r}^2\sigma_n(\bar{r})d\bar{r},
\end{equation}
where $m_n(r)$ and $\sigma_n(r)$ denote the quasi-local mass
function and energy density, respectively. The compactness function
is directly related to the formation of event horizon in the
considered spherically symmetric geometry. It is observed that the
compactness function must be less than unity to avoid the existence
of an event horizon. We consider the interior region of gravastar as
dS manifold with cosmological constant $\Lambda_i$ having constant
energy density, i.e., $\sigma_i=\Lambda_i/8\pi\geq0$. The respective
compactness function of the interior region is given as
$\lambda_{i}(r)=8\pi \sigma_{i}r^2/3=r^2/L^2_{i}$ with
$L_{i}=\sqrt{3/\Lambda_i}$, for $r<a(\tau)$ where $r=a(\tau)=a$ is
the position of gravastar shell and $\tau$ denotes the proper time.
It is noted that if $\lambda_i(r)\rightarrow 1$ then $r\rightarrow
(3/\Lambda_i)^{1/2}$ which represents the position of
DS/cosmological horizon.

The gravastars and regular BHs both are considered to overcome the
central singularity of the gravitational collapse. But, we are
interested to observe the possible outcomes of the inner flat/dS in
the presence of external regular BH geometry just like the
construction of thin-shell WHs from two equivalent copies of regular
BHs. Here, we use Bardeen-dS manifold as an exterior geometry with
constant energy density. The corresponding action is given as
(Fernando 2017)
\begin{eqnarray}\nonumber
S=\int d^4
x\sqrt{-g}\left[\frac{1}{16\pi}\left(R-2\Lambda_e\right)-\frac{1}{4\pi}\mathcal{L}(F)\right],
\end{eqnarray}
where $\Lambda_e$ is the cosmological constant and $R$ represents
the scalar curvature. Here, $\mathcal{L}(F)$ is a function of
Maxwell invariant $F=1/4F_{\mu\nu}F^{\mu\nu}$ defined as (Fernando
2017)
\begin{eqnarray}\nonumber
\mathcal{L}(F)=\frac{3m}{|Q|Q^2}\left(\frac{\sqrt{2Q^2F}}{1+\sqrt{2Q^2F}}\right)^\frac{5}{2}.
\end{eqnarray}
where $m$ is the mass of BH and $Q$ represents the magnetic monopole
charge of the geometry. The Maxwell invariant can be obtained as
$F=Q^2/2r^4$. Hence, we have
\begin{eqnarray}\nonumber
\mathcal{L}(F)=\frac{3mQ^2}{\left(r^2+Q^2\right)^\frac{5}{2}}.
\end{eqnarray}
The mass function of the exterior spacetime (Bardeen-dS BH) can be
expressed as (Fernando 2017)
\begin{equation}\label{3}
m_e(r)=\frac{\Lambda_e
r^3}{6}+\frac{mr^3}{(r^2+Q^2)^\frac{3}{2}},\quad r>a.
\end{equation}
Hence, the respective compactness function becomes
\begin{equation}\label{3}
\lambda_{e}(r)=\frac{2m_e(r)}{r}=2mr^2/(r^2+Q^2)^{3/2}+r^2/L^2_{e},\quad
r>a,
\end{equation}
with $L_{e}=\sqrt{3/\Lambda_e}$. This spacetime can further be
reduces to different manifolds for some specific values of the
physical parameters.
\begin{itemize}
\item If $\Lambda_e=0$, $Q>0$ and $M>0$, then it denotes the Bardeen
BH (Bardeen 1968).
\item If $\Lambda_e>0$, $Q=0$ and $M>0$, it represents the Schwarzschild-dS BH.
\item If $\Lambda_e=0=Q$ and $M>0$, then it corresponds to the Schwarzschild BH.
\item If $\Lambda_e>0$ and $Q=0=M$, then it is clearly dS spacetime.
\end{itemize}
To avoid the event horizon in the gravastar geometry, the following
conditions must be satisfied, i.e., (i) $\Lambda_i<3/r^2$ for $r<a$,
(ii) $\Lambda_e<3/r^2-6m/(r^2+Q^2)^{3/2}$ for $r>a$. These
conditions indicate that $\Lambda_i$ must be greater than
$\Lambda_e$ at $r=a$, i.e., $\Lambda_i>\Lambda_e$. This provides the
basic motivation behind the construction of the gravastar model that
the energy density of the surrounding region of the shell is less
than the inner region.

The inner and outer regions are connected at the timelike
hypersurface ($\Sigma$), i.e., a (2+1)-dimensional spacetime, also
referred to as a gravastar shell with radius $r=a$. Such matching is
followed by Visser's cut and paste procedure. This approach is
useful to avoid the presence of an event horizon as well as a
central singularity in the geometry of gravastars. We use Israel
junction conditions to ensure that the developed structure is a
solution to the field equations. These conditions are useful to
evaluate the mass of the shell and its potential function. The
continuity of the line element (first junction condition) shows that
the induced metric on $\Sigma$ is identical for the metrics on both
sides of $\Sigma$, i.e., $ds^2_i=ds^2_e=ds^2_\Sigma$ implying that
$k_{i}(1-\lambda_i(a))=k_e(1-\lambda_e(a))$. For the choice of
Bardeen-dS BH, $k_e\rightarrow 1$ and hence
$k_{i}\rightarrow(1-\lambda_e(a))/(1-\lambda_i(a))$ that re-scales
the time coordinate in the interior region.

The respective coordinates of $\Sigma$ are denoted as
$\xi^{a}=(\tau,\theta,\phi)$ and the line element (\ref{1}) at $r=a$
becomes
\begin{equation}\label{4}
ds^2=\left[-(1-\lambda(a))+(1-\lambda(a))^{-1}(da/d\tau)^{2}
(d\tau/dt)^{2}\right]dt^2+a^2d\Omega^2.
\end{equation}
The corresponding inducing metric at $\Sigma$ is given as
\begin{equation}\label{5}
ds^2_\Sigma=h_{ab}d\xi^a d\xi^b=-d\tau^{2}+a^2d\Omega^2,
\end{equation}
where $h_{ab}=g_{\mu\nu}e^{\mu}_ae^{\nu}_b$ is the induced metric
tensor of $\Sigma$ with $e^{\mu}_a=\partial x^\mu/\partial \xi^a$.
By comparing Eqs.(\ref{4}) and (\ref{5}), we obtain
\begin{eqnarray}\label{6}
dt/d\tau=\dot{t}=\begin{cases}(1-\lambda_i(a)+\dot{a}^2)^{1/2}/1
-\lambda_i(a),\quad \text{for the interior region}\\
(1-\lambda_e(a)+\dot{a}^2)^{1/2}/1-\lambda_e(a), \quad \text{for the
exterior region}\end{cases}
\end{eqnarray}
The second condition leads to the continuity of extrinsic curvature
($K^{a}_{b}$) that shows smooth connection between inner and outer
regions at $\Sigma$, i.e.,
$[K^{a}_{b}]=K^{a}_{b(i)}-K^{a}_{b(e)}=0$. If $[K^{a}_{b}]=0$, then
it represents a boundary surface between these spacetimes, otherwise
it leads to a thin-shell. The respective non-vanishing components of
$K^{a}_{b}$ at shell radius $r=a$ for Eq.(\ref{1}) are given as
\begin{eqnarray}\label{7}
K^{\tau}_\tau&=&\begin{cases}(\lambda_i'(a)+2\ddot{a}^2)/(1-\lambda_i(a)
+\dot{a}^2)^{1/2},\quad \text{for the interior region}\\
(\lambda_e'(a)+2\ddot{a}^2)/(1-\lambda_e(a) +\dot{a}^2)^{1/2},\quad
\text{for the exterior region}\end{cases} \\\label{8}
K^{\theta}_\theta&=&\begin{cases}(1-\lambda_i(a)
+\dot{a}^2)^{1/2}/a,\quad \text{for the interior region}\\
(1-\lambda_e(a) +\dot{a}^2)^{1/2}/a,\quad \text{for the exterior
region}\end{cases} \\\label{9}
K^{\phi}_\phi&=&K^{\theta}_\theta\sin^2\theta,
\end{eqnarray}
where $\lambda'(a)$ denotes differentiation of $\lambda(a)$ with
respect to $a$. The matter surfaced at shell produces discontinuity
in the extrinsic curvatures of both spacetimes. The standard
expression of the stress-energy tensor for timelike hypersurface can
be expressed as
\begin{equation}\label{10}
S^{a}_{b}=-\frac{1}{8\pi}\{[K^{a}_{b}]-\delta^{a}_{b}K\},
\end{equation}
where $K=[K^{a}_{a}]$ and hence, we have (Lake 1979)
\begin{equation}\label{11}
-[K^{\theta}_\theta]/4\pi=(1-\lambda_e(a) +\dot{a}^2)^{1/2}/4\pi
a-(1-\lambda_i(a) +\dot{a}^2)^{1/2}/4\pi a=\rho,
\end{equation}
here $\rho=M/4\pi a^2$ is the energy density and $M$ denotes the
mass of gravastar shell. Using the values of compactness function of
both spacetimes, the above equation can be written as
\begin{equation}\label{12}
M+a\left[1-\frac{2ma^2}{(a^2+Q^2)^{3/2}}-\frac{a^2}{L^2_{e}}
+\dot{a}^2\right]^{1/2}-a\left[1-\frac{a^2}{L^2_{i}}
+\dot{a}^2\right]^{1/2}=0,
\end{equation}

The effective potential of gravastar shell is followed from the
equation
\begin{eqnarray}\label{13}
V(m,Q,M,L_i,L_e)+\dot{a}^2/2=0.
\end{eqnarray}
Now, we solve Eq.(\ref{12}) for $\dot{a}^2$ to determine the
mathematical expression of the effective potential as
\begin{eqnarray}\nonumber
V(a,m,Q,M,L_i,L_e)&=&\frac{1}{2}-\frac{a^6}{8 L_i^4 M^2}+\frac{m
a^6}{2 L_i^2 M^2 \left(Q^2+a^2\right)^{3/2}}+\frac{a^6}{4 L_i^2 M^2
L_e^2}\\\nonumber&-&\frac{a^2}{4 L_i^2}-\frac{a^6}{8 M^2
L_e^4}-\frac{m a^6}{2 M^2 L_e^2 \left(Q^2+a^2\right)
^{3/2}}-\frac{M^2}{8 a^2}\\\label{14} &-&\frac{m^2 a^6}{2 M^2
\left(Q^2+a^2\right)^3}-\frac{m a^2}{2
\left(Q^2+a^2\right)^{3/2}}-\frac{a^2}{4 L_e^2}.
\end{eqnarray}
We assume that the matter surface at gravastar shell follows the EoS
(Rocha et al. 2008a; 2008b)
\begin{eqnarray}\label{15}
p=(1-\beta)\rho,
\end{eqnarray}
here $p$ is the surface pressure and $\beta$ is a dimensionless
constant. It is interesting to mention here that different values of
$\beta$ represent different types of matter distribution, i.e.,
standard energy ($\beta\leq1.5$), dark energy ($1.5<\beta\leq2$) and
phantom energy ($\beta>2$). The continuity of perfect fluid gives
the relationship between the surface stresses of gravastar as
\begin{eqnarray}\label{16}
4\pi\frac{d}{d\tau}(a^2\rho)+4\pi p\frac{da^2}{d\tau}=0,
\end{eqnarray}
which can be expressed as
\begin{eqnarray}\label{17}
\frac{d\rho}{da}=-\frac{2}{a}(\rho+p).
\end{eqnarray}
Using Eq.(\ref{15}) and $\rho=M/4\pi a^2$ in (\ref{17}), then
solving, it follows that
\begin{eqnarray}\label{18}
M=ha^{2(\beta-1)},
\end{eqnarray}
where $h$ is an integrating constant.

The general expression of potential function is obtained by using
Eq.(\ref{18}) in (\ref{14}) as
\begin{eqnarray}\nonumber
V(a,m,Q,L_i,L_e,h,\beta)&=&\frac{a^{10-4 \beta}}{4 h^2 L_i^2
L_e^2}-\frac{a^{10-4 \beta}}{8 h^2 L_i^4}+\frac{m a^{10-4 \beta}}{2
h^2 L_i^2 \left(Q^2+a^2\right)^{3/2}}-\frac{a^2}{4
L_e^2}\\\nonumber&-&\frac{m^2 a^{10-4 \beta}}{2 h^2
\left(Q^2+a^2\right)^3}-\frac{m a^{10-4 \beta}}{2 h^2 L_e^2
\left(Q^2+a^2\right)^{3/2}}-\frac{1}{8} h^2 a^{4
\beta-6}\\\label{19}&-&\frac{a^2}{4 L_i^2}-\frac{m a^2}{2
\left(Q^2+a^2\right)^{3/2}}-\frac{a^{10-4 \beta}}{8 h^2
L_e^4}+\frac{1}{2}.
\end{eqnarray}
To overcome the dependence of the potential function on the
integration constant $h$, we redefine the physical parameters as
follows
\begin{eqnarray}\label{20}
m\rightarrow m h^{\frac{1}{3-2\beta}},\quad Q\rightarrow Q
h^{\frac{1}{3-2\beta}},\quad L_i\rightarrow L_i
h^{\frac{1}{3-2\beta}},\quad L_e\rightarrow L_e
h^{\frac{1}{3-2\beta}},
\end{eqnarray}
so that
\begin{eqnarray}\nonumber
V(a,m,Q,L_i,L_e,\beta)&=&\frac{m a^{10-4 \beta}}{2 L_i^2
\left(Q^2+a^2\right)^{3/2}}-\frac{m^2 a^{10-4 \beta}}{2
\left(Q^2+a^2\right)^3}-\frac{m a^2}{2
\left(Q^2+a^2\right)^{3/2}}\\\nonumber&-&\frac{m a^{10-4 \beta}}{2
L_e^2 \left(Q^2+a^2\right)^{3/2}}-\frac{a^{10-4 \beta}}{8
L_i^4}+\frac{a^{10-4 \beta}}{4 L_i^2 L_e^2}-\frac{a^2}{4
L_i^2}-\frac{a^2}{4 L_e^2}\\\label{21}&-&\frac{1}{8} a^{4
\beta-6}-\frac{a^{10-4 \beta}}{8 L_e^4}+\frac{1}{2}.
\end{eqnarray}
This is used to explore the expanding as well as collapsing
characteristics of gravastar shell surrounded by the regular BH for
suitable values of physical parameters $m$, $Q$, $L_i$, $L_e$ and
$\beta$. Here, we consider the following physical parameters with
their units, i.e., $[a]=[m]=[Q]=[\Lambda]=L$, $[\rho]=[p]=1/L$ and
$[\Lambda_i]=[\Lambda_e]=1/L^2$. The values of all parameters used
in this paper have frequently been used in literature that explores
the possible existence of bounded excursion gravastar as well as
their stable configuration (Rocha et al 2008a; 2008b; Chan et al
2009a; 2009b; 2010).

\section{Gravitational Collapse and Gravastars}

This section explores the outcomes of the developed structure that
can be a BH, stable/unstable, or ``bounded excursion" gravastar or a
Minkowski or dS geometry for various matter distributions at
hypersurface. The potential function of the gravatar shell is used
to characterize the final fate of the collapse through suitable
physical parameters. The effective potential of the gravastar shell
is followed from Eq.(\ref{13}). This expression is also referred to
as an equation of motion of the shell in one dimension after the
radial perturbation. We have noted that the exact solution of the
equation of motion (\ref{13}) is not possible for the shell radius
after radial perturbation as it is a non-linear equation. Thus we
observe the shell's motion through a linearized form of this
equation without going through the complete solution after the
perturbation. For this purpose, we expand the potential function
about equilibrium shell radius by using Taylor series expansion up
to second-order terms as
\begin{equation}\nonumber
V(a)=V(a_0)+\frac{dV}{da}|_{a=a_0}
(a-a_0)+\frac{d^2V}{da^2}|_{a=a_0}\frac{(a-a_0)^2}{2}+O((a-a_0)^3),
\end{equation}
where $V(a_0)=0=\frac{dV} {da}|_{a=a_0}$. By considering $x=a-a_0$,
we obtain
\begin{equation}\label{27a}
\dot{x}^2+\omega^2x^2\simeq 0,
\end{equation}
where
\begin{equation}\nonumber
\omega^2=\frac{d^2V(a)}{da^2}|_{a=a_0}.
\end{equation}
Differentiating Eq.(\ref{27a}) with respect to proper time, it
follows that
\begin{equation}\label{29a}
\ddot{x}+\omega^2x\simeq 0.
\end{equation}
\begin{figure}\centering\label{3}
\epsfig{file=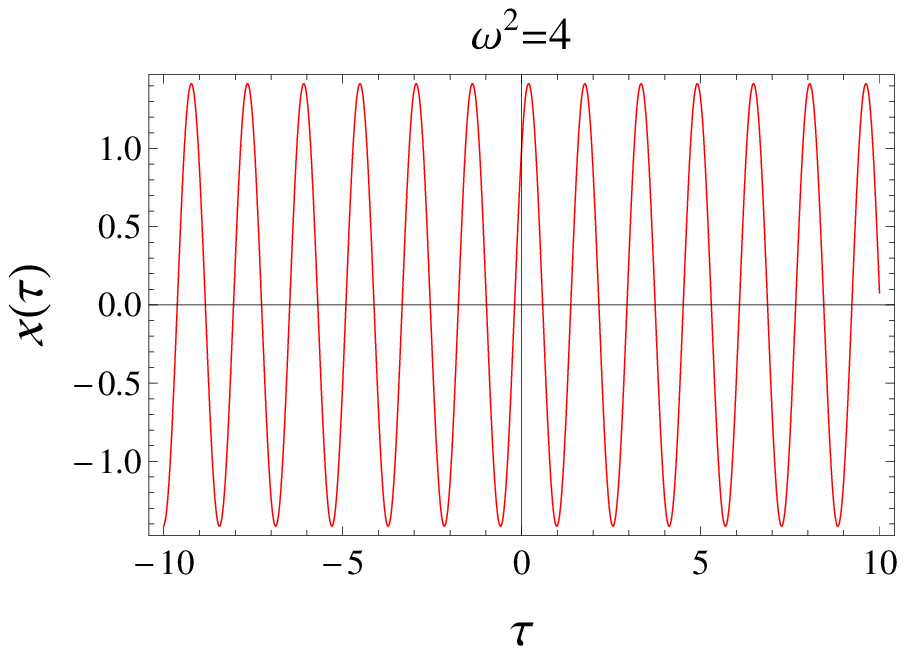,width=.5\linewidth}\epsfig{file=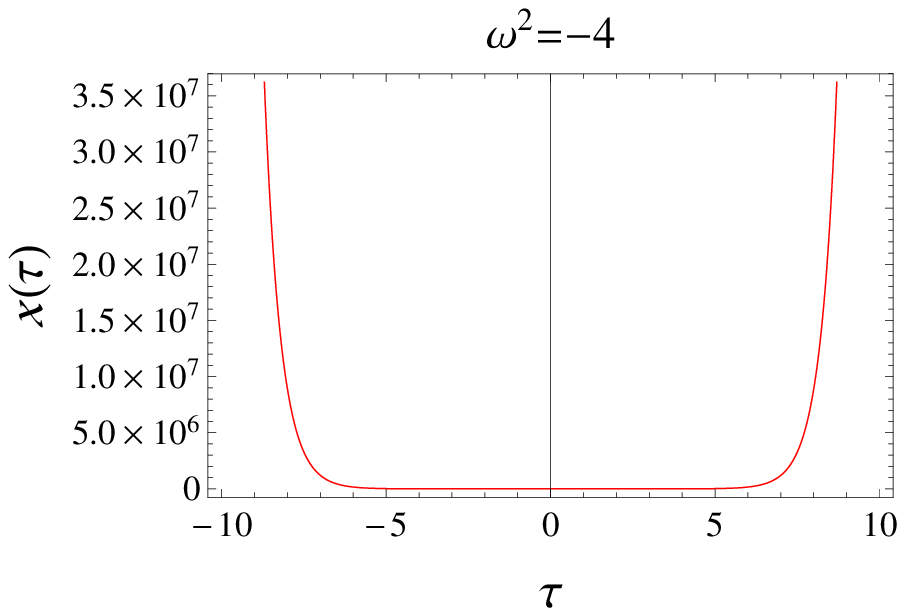,width=.5\linewidth}
\caption{The graphical behavior of the solution of Eq.(\ref{29a})
with different values of $\omega^2$.}
\end{figure}

This equation shows stable and unstable configurations of a
thin-shell that depends on $\omega^2$. It is found that thin-shell
expresses oscillation about $x=0$ for $\omega^2>0$ as shown in the
left plot of Figure \textbf{1}. Hence, the developed structure
indicates oscillation about the equilibrium shell radius
($x=0\Rightarrow a=a_0$) and remains stable. This leads to a stable
configuration of a thin-shell. We can write this condition as
\begin{equation}\label{30a}
\frac{d^2V(a)}{da^2}|_{a=a_0}>0.
\end{equation}
If $\omega^2<0$, then the shell radius represents the exponential
behavior which corresponds to the unstable behavior as shown in the
right plot of Figure \textbf{1}. The respective condition for the
unstable configuration can be expressed as
\begin{equation}\label{31a}
\frac{d^2V(a)}{da^2}|_{a=a_0}<0.
\end{equation}
Hence, the developed structure shows stable configuration against
the radial perturbation for an equilibrium shell radius $a=a_0$, if
\begin{eqnarray}\label{22}
V(a_0)=0, \quad V'(a_0)=0, \quad V''(a_0)>0.
\end{eqnarray}
The shell's motion along the radial direction vanishes at
equilibrium shell radius, i.e., $\dot{a}_{0}=0=\ddot{a}_{0}$. If
$V''(a_{0})<0$, then it shows unstable behavior while it is
unpredictable when $V''(a_{0})=0$.

We would like to mention here that we have explored the stability of
thin-shell gravastar in the background of interior dS and exterior
regular BHs (Bardeen and Bardeen-dS BHs) (Sharif and Javed 2020). We
have investigated the physical viability of the developed model
through the energy conditions and explored its stability by using
radial perturbation about the equilibrium shell radius. We have
found that thin-shell gravastars show large stable regions for the
Bardeen-dS BH as compared to the Bardeen BH. We have concluded that
stable regions exist near the formation of the expected event
horizon (Sharif and Javed 2020). The present manuscript is devoted
to exploring the outcomes of the developed structure filled with
stiff and dust fluid distributions. Here, we extend this work for
the possible existence of bounded excursion gravastar with different
choices of interior and exterior geometries. The stable bounded
excursion is a less stringent concept to determine the stability of
a geometrical structure. According to this notion, the shell's
motion must be bounded in between an interval $(a_1,a_2)$ with
$a_1<a_2$ and follows the conditions
\begin{eqnarray}\label{23}
V(a_1)=0, \quad V'(a_1)\leq0, \quad V(a_2)=0, \quad V'(a_2)\geq0,
\end{eqnarray}
with $V(a)<0$, $\forall a\in(a_1,a_2)$.

We use this approach to discuss the outcomes of the geometrical
structure by using the potential function and its derivative with
respect to the shell radius. For this purpose, we solve
simultaneously the following equations $V(m_c, Q_c, L_{ic}, L_{ec},
a_c)=0=V'(m_c,Q_c, L_{ic},L_{ec}, a_c)$ and evaluate the respective
critical values of the physical parameters like $m_c$, $Q_c$,
$L_{ic}$, $L_{ec}$ and $a_{c}$. The critical value of the shell
radius ($a_c$) demonstrates the position of the shell at which it
has neither expanding nor collapsing nature. These points of the
geometrical structure can also be referred to as the saddle points.
Similarly, the remaining critical values of the physical parameters
also explain the position of the shell at saddle points. Due to
complicated nature of Eq.(\ref{21}), we cannot manipulate it
analytically and hence consider some special cases. The specific
values of physical parameters lead to different interior and
exterior geometries of the shell. We do not consider the case
$m=\Lambda_i=\Lambda_e=0$ as it corresponds to the Minkowski
spacetime in both interior as well as exterior regions and hence
thin-shell disappears. Following special cases recover the previous
results (Rocha et al 2008a; 2008b; Chan et al 2009b; 2010).
\begin{itemize}
\item $m=0=\Lambda_e$ and $\Lambda_i\neq0$.
\item $m=0=\Lambda_i$ and $\Lambda_e\neq0$.
\item $m\neq0$, $Q=0$ and $\Lambda_i=0=\Lambda_e$.
\item $m\neq0$, $Q=0$ and $\Lambda_i\neq0\neq\Lambda_e$.
\end{itemize}
Here, we consider the following cases
\begin{itemize}
\item $m\neq0\neq Q$ and $\Lambda_i=0=\Lambda_e$.
\item $m\neq0\neq Q$, $\Lambda_i\neq0$ and $\Lambda_e=0$.
\item $m\neq0\neq Q$ and $\Lambda_i\neq0\neq\Lambda_e$.
\end{itemize}
In the following, we explore a suitable case for the construction of
bounded excursion gravastar through effective potential with two
different choices of matter distribution at the shell, i.e., stiff
($\beta=0$) and dust ($\beta=1$) fluids.

\subsection{ Stiff Fluid Shell}

Here, we assume that the shell of the constructed model is filled
with stiff fluid and the effective potential for such a matter
distribution can be obtained by using $\beta=0$ in Eq.(\ref{21}).
The corresponding expressions of the potential function and its
derivative to $a$ is given as
\begin{eqnarray}\nonumber
V(a)&=&\frac{1}{2}+\frac{a^{10}}{4 L_e^2 L_i^2}-\frac{a^{10}}{8
L_e^4}-\frac{a^{10}}{8 L_i^4}-\frac{1}{8 a^6}-\frac{a^2}{4
L_e^2}-\frac{a^2 m}{2 \left(a^2+Q^2\right)^{3/2}}-\frac{a^2}{4
L_i^2}\\\label{24}&-&\frac{a^{10} m}{2 \left(a^2+Q^2\right)^{3/2}
L_e^2}+\frac{a^{10} m}{2 \left(a^2+Q^2\right)^{3/2}
L_i^2}-\frac{a^{10} m^2}{2 \left(a^2+Q^2\right)^3},\\\nonumber
V'(a)&=&\frac{5 a^9}{2 L_e^2 L_i^2}-\frac{5 a^9}{4 L_e^4}-\frac{a
m}{\left(a^2+Q^2\right)^{3/2}}+\frac{3 a^{11} m}{2
\left(a^2+Q^2\right)^{5/2} L_e^2}-\frac{5 a^9}{4 L_i^4}-\frac{a}{2
L_i^2}\\\nonumber&+&\frac{3}{4 a^7}-\frac{3 a^{11} m}{2
\left(a^2+Q^2\right)^{5/2} L_i^2}+\frac{3 a^{11}
m^2}{\left(a^2+Q^2\right)^4}-\frac{5 a^9
m}{\left(a^2+Q^2\right)^{3/2} L_e^2}\\\label{25}&+&\frac{5 a^9
m}{\left(a^2+Q^2\right)^{3/2} L_i^2}-\frac{5 a^9
m^2}{\left(a^2+Q^2\right)^3}-\frac{a}{2 L_e^2}+\frac{3 a^3 m}{2
\left(a^2+Q^2\right)^{5/2}},
\end{eqnarray}
respectively. These are used to discuss the dynamical configuration
of a shell with suitable values of the physical parameters. We solve
simultaneously $V(a)=0$ and $V'(a)=0$ for suitable values of $Q$.
The solutions of this system are used to determine the shell's
radius at which the shell neither collapses nor expands. These
values are also referred to as critical values of the physical
parameters. We have obtained the respective critical values of
physical parameters like $a_c$, $Q_c$, $m_{c}$, $L_{ic}$ and
$L_{ec}$ for stiff matter distribution at the constructed shell as
shown in Table {\textbf{1}}. This table also explains the outcomes
of the developed structure with different choices of interior and
exterior geometries. We explore the graphical behavior of the
potential function corresponding to critical values of physical
parameters. Here, we use three different possibilities to explore
the stable configuration of the developed structure for the critical
values, i.e., $m_1<m_c<m_2$, $Q_1<Q_c<Q_2$, $L_{i1}<L_{ic}<L_{i2}$
and $L_{e1}<L_{ec}<L_{e2}$. To obtain suitable results, we compare
the graphical behavior through these choices, i.e., the system
behavior for critical values and other values that are greater or
smaller than critical values. For this purpose, we assume suitable
values of the physical parameters close to the critical ones.

\subsubsection*{Case (i): $m\neq0\neq Q$ and
$\Lambda_i=0=\Lambda_e$}

This case denotes the internal flat region with external Bardeen BH
and the respective critical values of physical parameters are shown
in Table \textbf{1}. For these values, the potential function shows
neither expansion nor collapse at $a=a_c$ as shown in Figures
\textbf{2-4}. It is found that the effective potential approaches to
$-\infty$ as $a\rightarrow \infty$ or $0$. Thus the strictly
negative behavior of potential function shows the formation of BH as
an outcome of the gravitational collapse of a star. We also analyze
the effects of mass and charge on the dynamical configuration of the
star. Here, we obtain the results for different choices of mass and
charge by considering $m<m_c$, $m=m_c$, $m>m_c$ and $Q<Q_c$,
$Q=Q_c$, $Q>Q_c$ as shown in the left and right plots of Figures
\textbf{2-4}, respectively.
\begin{table}[ht]
\caption{Critical values of physical parameters and outcomes of the
developed structure for a stiff fluid shell with different choices
of internal and external geometries. Here, we use the internal
region (IR), external region (ER), structure (St), figures (Fig.)
flat region (FR), Bardeen BH (B), Bardeen-dS BH (BdS), the normal
star (NS), and gravastar (GS).} \label{table:1} \vspace{0.5cm}
\centering
\begin{tabular}{ |c c c c c c c c c |c| }
\hline \multicolumn{9}{|c|}{{Final outcomes of the developed
structure for stiff fluid shell ($\beta=0$)}}\\ \hline
$Q_c$ & $a_c$ & $m_c$ & $L_{ic}$ & $L_{ec}$ & IR & ER & St & Fig \\
\hline\hline
0.4 & 1.041184 & 0.631676 & - & - & FR & B & BH & \textbf{2}\\
0.5 & 1.026782 & 0.702327 & - & - & FR & B & BH & \textbf{3} \\
2.5 & 0.937496 & 10.33415 & - & - & FR & B & BH & \textbf{4} \\
\hline
0.4 & 1.041184 & 0.422023 & 2.944657 & - & dS & B & GS & \textbf{5}\\
0.5 & 1.026782 & 0.435288 & 2.574298 & - & dS & B & GS & \textbf{6} \\
2.5 & 0.937496 & 10.33415 & 3.8117$\times 10^7$ & - & DS & B & GS & \textbf{7} \\
\hline
0.4 & 1.041184 & 0.549544 & 2.944657 & 2.793761 & dS & BdS & GS & \textbf{8}\\
0.5 & 1.026782 & 0.572372 & 2.574298 & 2.357312 & dS & BdS & GS & \textbf{9} \\
2.5 & 0.937496 & 10.33415 & 3.8117$\times 10^7$ & 1.67201$\times 10^7$ & dS & BdS & GS & \textbf{10} \\
\hline
\end{tabular}
\end{table}

For $m<m_c$, the potential function has two real roots say $a_1$ and
$a_2$ with $a_1<a_2$. It is found that $V(a)<0$ for $a<a_1$,
$V(a)>0$ for $a_2<a<a_1$ and $V(a)<0$ for $a>a_2$. This represents
the unstable structure under small perturbation which leads to star
either collapses until $a=0$ and leaves behind a Bardeen BH or
expands forever to form a flat spacetime (left plot of Figure
\textbf{2}). For $m=m_c$, the star shows collapsing behavior for
$a<a_c$ until $a\rightarrow0$ that leads to Minkowski spacetime and
stops collapse or expansion at $a=a_c$ and then collapses
continuously for $a>a_c$. The potential function is strictly
negative for $m>m_c$ which corresponds to the formation of BH as an
output of the collapse of a star as shown in Figure \textbf{2}. We
have the similar behavior of potential function for different values
of charge (Figures \textbf{2-4}). Hence, this case gives unstable
stable structure under small perturbation for every choice of the
physical parameters.
\begin{figure}\centering\label{2}
\epsfig{file=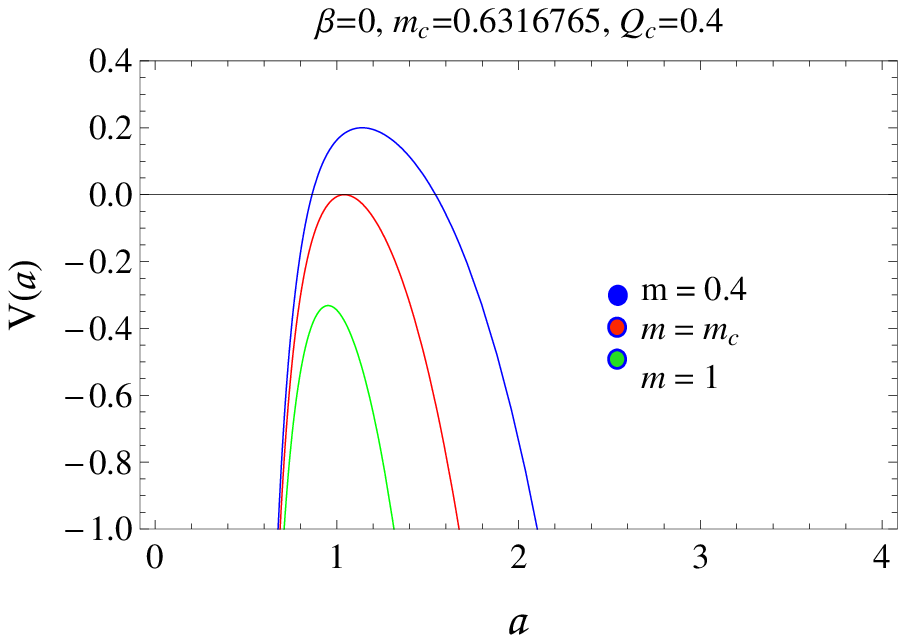,width=.5\linewidth}\epsfig{file=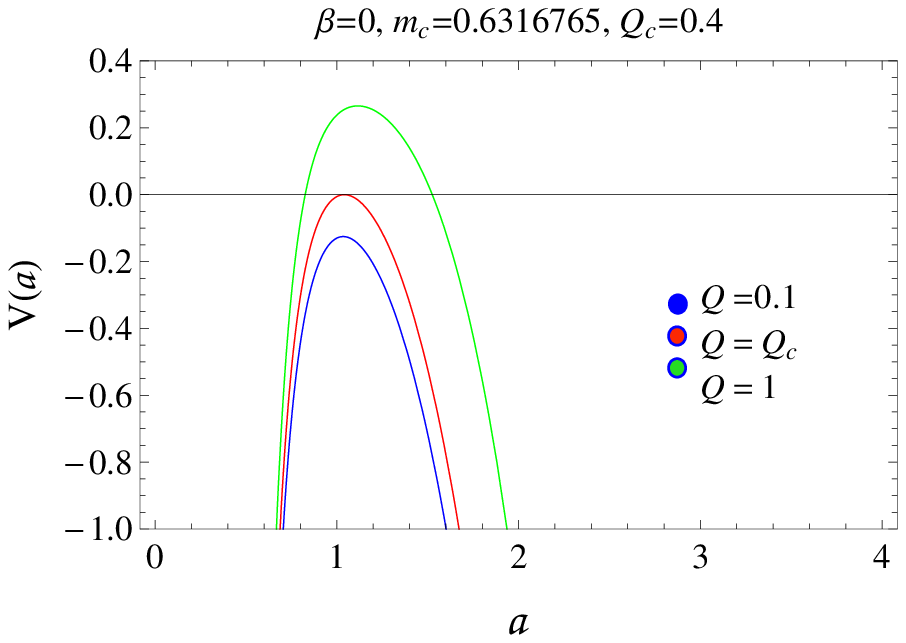,width=.5\linewidth}
\caption{The potential function for interior flat and exterior
Bardeen BH with different values of mass (left plot) and charge
(right plot).}
\epsfig{file=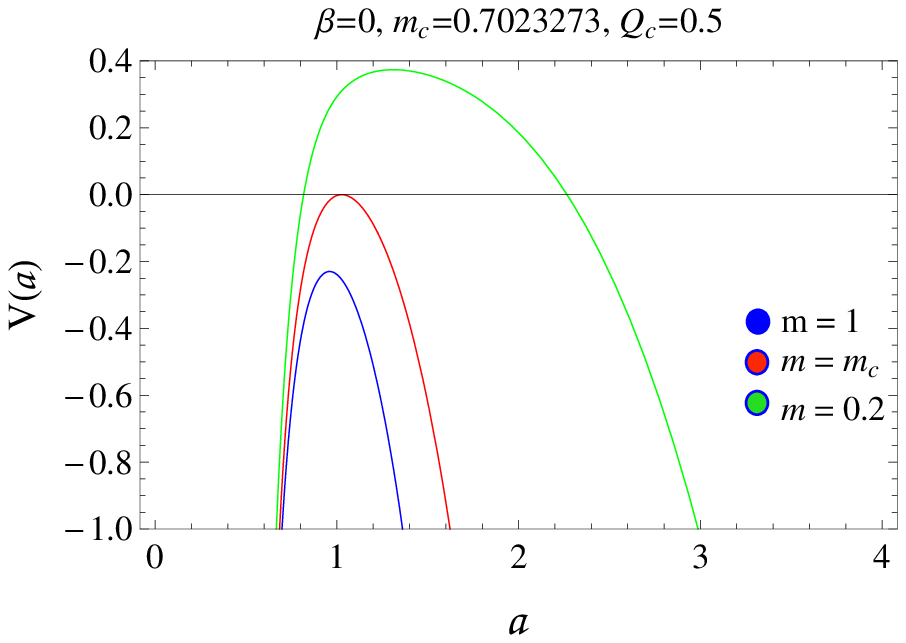,width=.5\linewidth}\epsfig{file=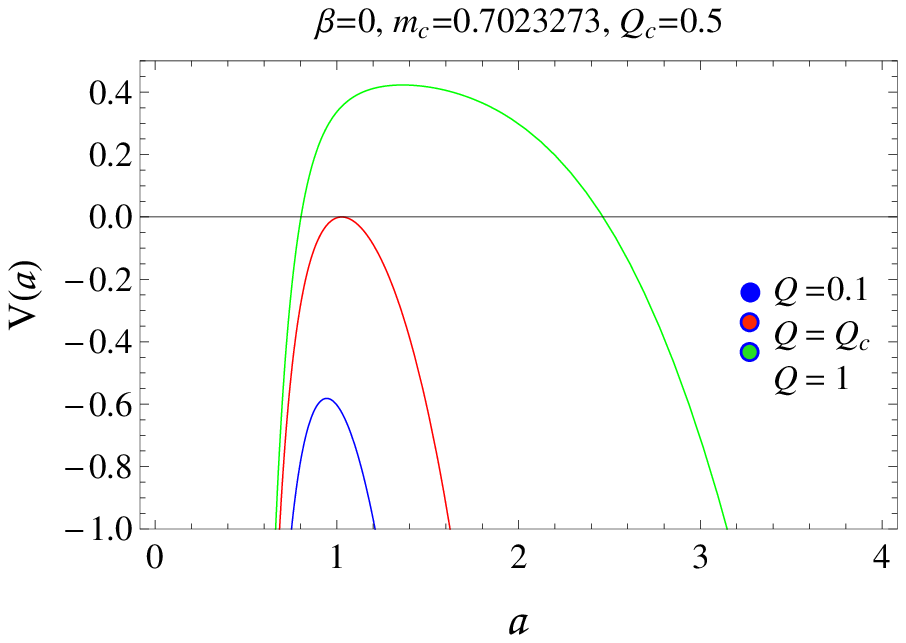,width=.5\linewidth}
\caption{The potential function for interior flat and exterior
Bardeen BH with different values of mass (left plot) and charge
(right plot).}
\epsfig{file=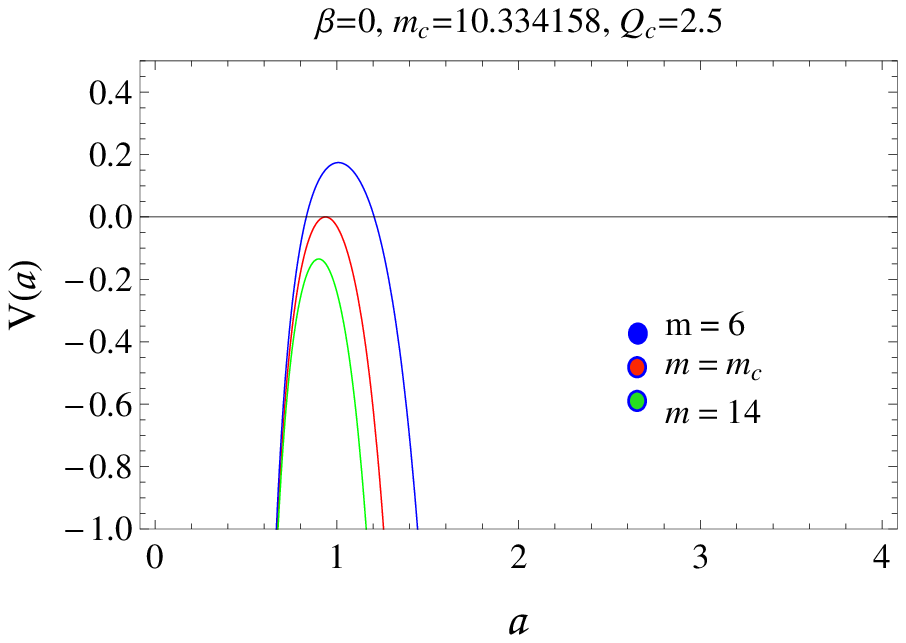,width=.5\linewidth}\epsfig{file=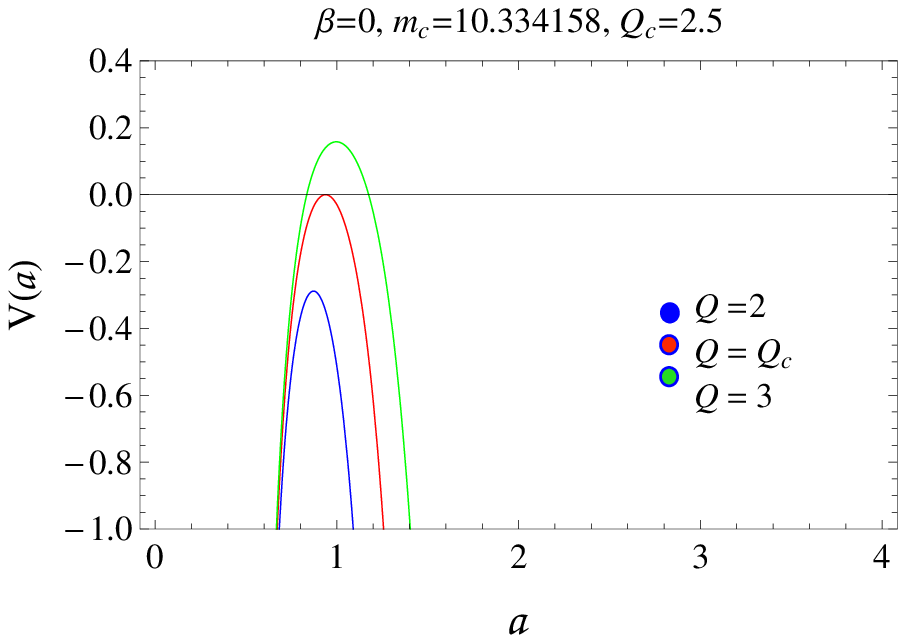,width=.5\linewidth}
\caption{The potential function for interior flat and exterior
Bardeen BH with different values of mass (left plot) and charge
(right plot).}
\end{figure}

\subsubsection*{Case (ii): $m\neq0\neq Q$, $\Lambda_i\neq0$ and
$\Lambda_e=0$}

This case represents the internal dS region with external Bardeen
BH. The corresponding critical values of physical parameters are
given in Table \textbf{1}. In this case, we observe the effects of
mass and $L_i$ on the final configuration of the developed structure
as shown in Figures \textbf{5-7}. This case represents the
possibility of the existence of bounded excursion gravastar for a
suitable choice of physical parameters. It is found that the star
shows collapsing behavior for critical values of the physical
parameters as given in Table \textbf{1}. The potential function has
four real roots for specific choices of $m$ and $L_i$ with
$0<a_1<a_2<a_3<a_4$. If $a<a_1$, then the potential function
approaches to $-\infty$ as $a\rightarrow0$ that represents the flat
spacetime. If $a_1<a<a_2$, then $V(a)>0$ which is unstable under
small perturbation. If $a_2<a<a_3$, then $V(a)<0$ implying the
existence of stable bounded excursion gravastar. If $a_3<a<a_4$,
then again $V(a)>0$ which shows unstable configuration under small
perturbation. The potential function again approaches $-\infty$ as
$a\rightarrow\infty$ for $a>a_4$ which leads to the formation of BH
(Figures \textbf{5-7}). It is found that the region of the existence
of gravastar is very small as compared to the BHs. Hence, the
dynamical model cannot completely exclude the possibility of the
formation of gravastar and BH structures. If the gravastar model
exists then there is also a possibility for the existence of BH and
vice versa. It is also interesting to mention here that the regions
of stable bounded excursion gravastar are enhanced by increasing
mass and $L_i$.
\begin{figure}\centering\label{2}
\epsfig{file=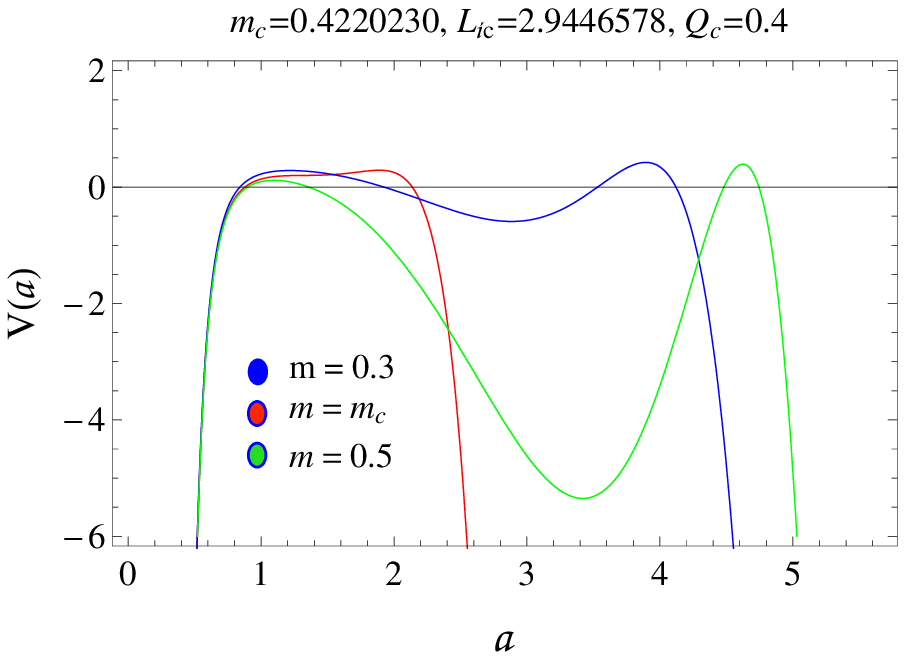,width=.5\linewidth}\epsfig{file=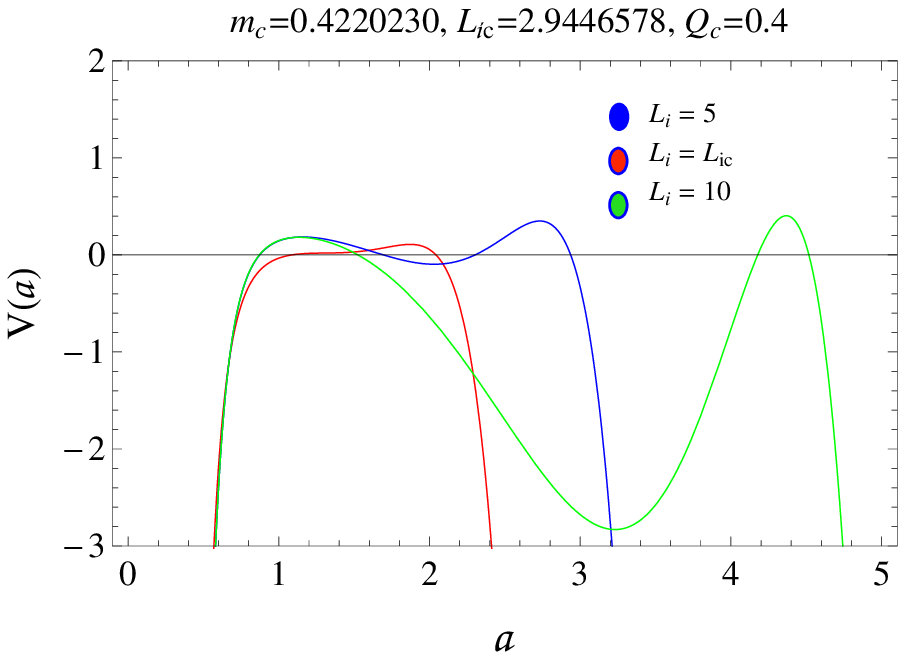,width=.5\linewidth}
\caption{The potential function for interior dS spacetime and
exterior Bardeen BH with different values of mass (left plot) and
$L_i$ (right plot).}
\epsfig{file=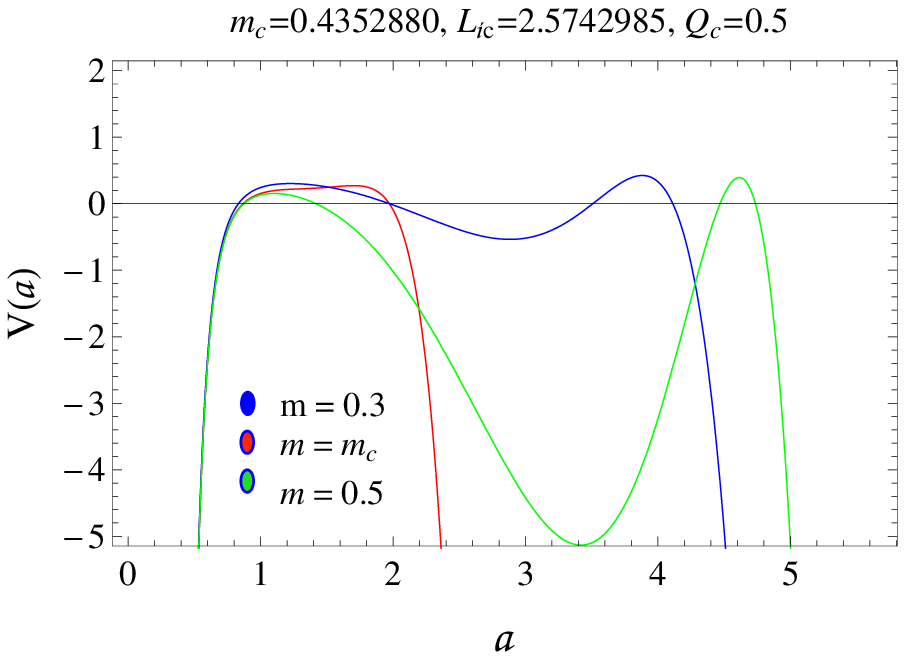,width=.5\linewidth}\epsfig{file=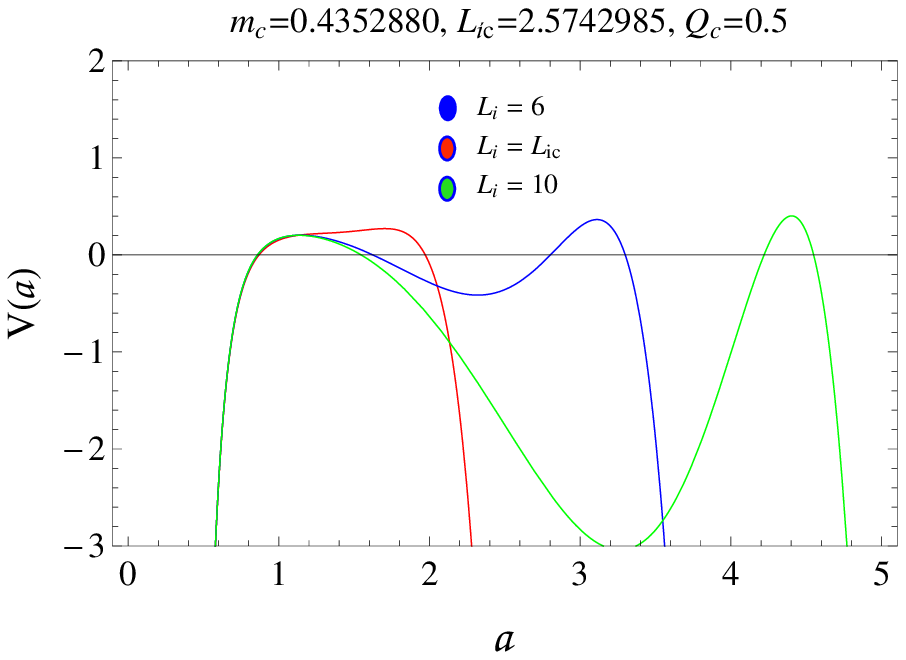,width=.5\linewidth}
\caption{The potential function for interior dS spacetime and
exterior Bardeen BH with different values of mass (left plot) and
$L_i$ (right plot).}
\epsfig{file=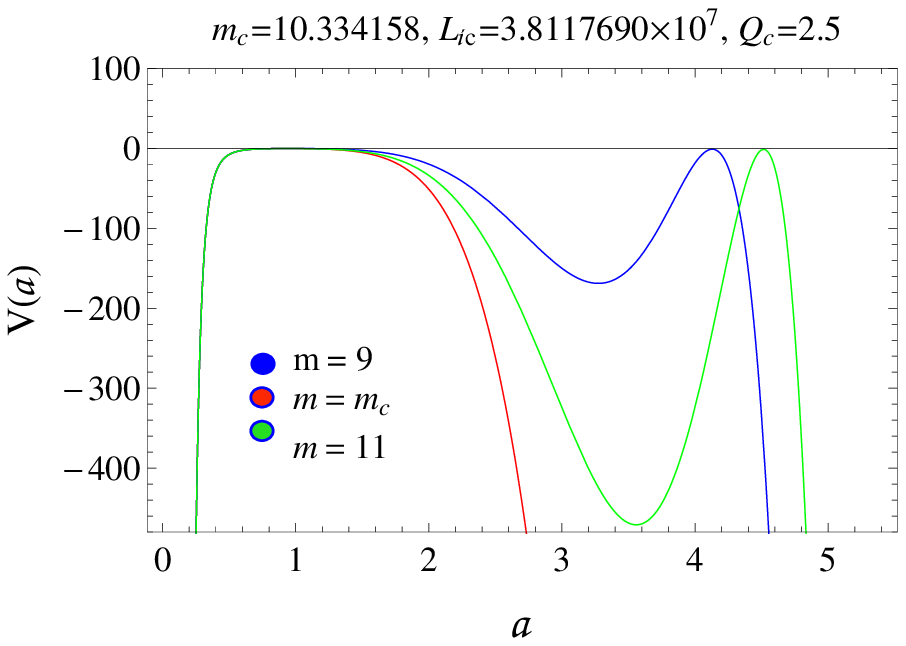,width=.5\linewidth}\epsfig{file=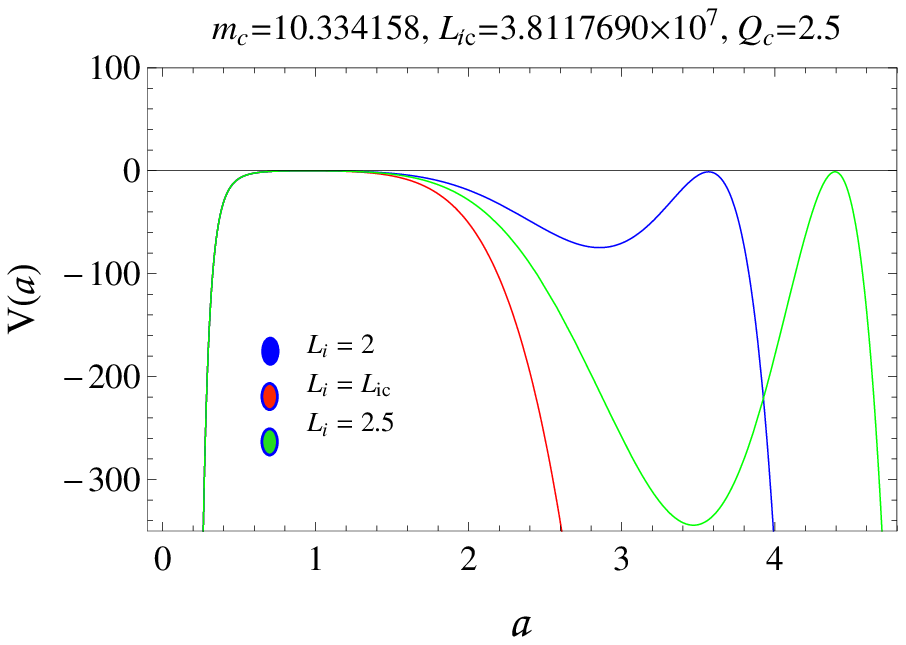,width=.5\linewidth}
\caption{The potential function for interior dS spacetime and
exterior Bardeen BH with different values of mass (left plot) and
$L_i$ (right plot).}
\end{figure}

\subsubsection*{Case (iii): $m\neq0\neq Q$ and $\Lambda_i\neq0\neq\Lambda_e$}

This is the general case that leads to the internal dS region and
external Bardeen-dS BH. The critical values of physical parameters
are given in Table \textbf{1}. To avoid the presence of an event
horizon in the developed structure, it is found that
$\Lambda_i>\Lambda_e$ which corresponds to $L_e>L_i$. It is observed
that the obtained critical values of $L_i$ and $L_e$ do not satisfy
the required inequality ($L_e>L_i$). Hence, the behavior of stars
must be collapsing for these choices of physical parameters. The
graphical behavior of potential function shows that the developed
structure represents collapse leading to BH if $L_e<L_i$ (Figures
\textbf{8-10}). Therefore, we assume the particular choice of $L_i$
and $L_e$ with $L_e>L_i$ for different values of $L_e$ and $Q$. It
is found that the developed structure shows the existence of stable
bounded excursion gravastar if $L_e>L_i$. The regions of stable
bounded excursion gravastar increase by increasing $L_e$ and $Q$.
\begin{figure}\centering\label{2}
\epsfig{file=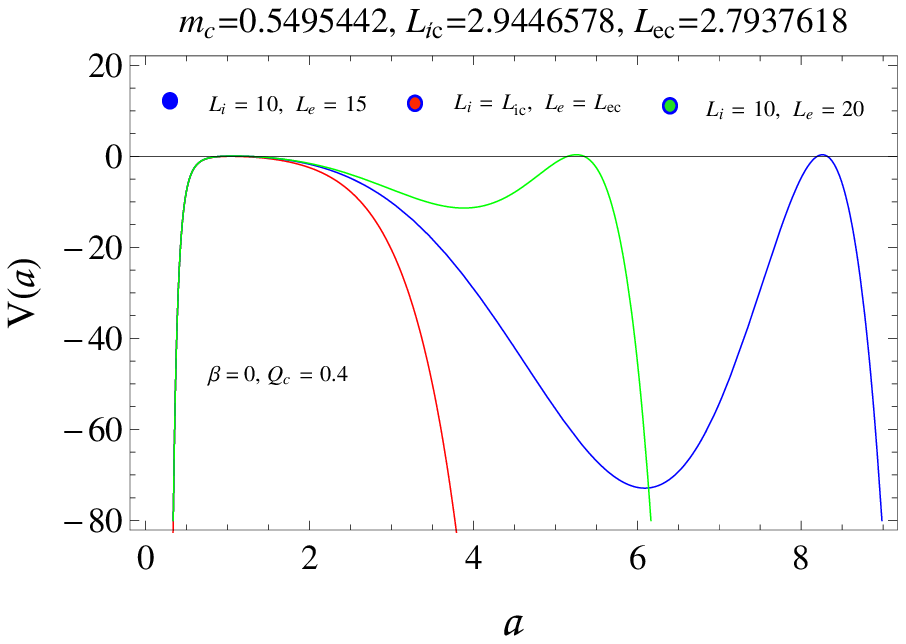,width=.5\linewidth}\epsfig{file=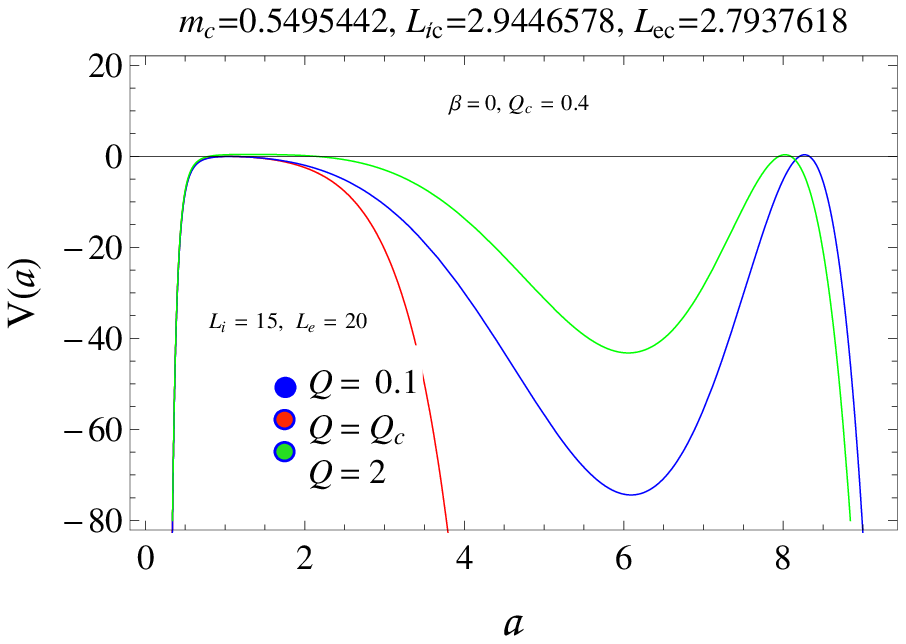,width=.5\linewidth}
\caption{The potential function for interior dS spacetime and
exterior Bardeen-dS BH with different values of $L_i$, $L_e$ (left
plot) and charge (right plot).}
\epsfig{file=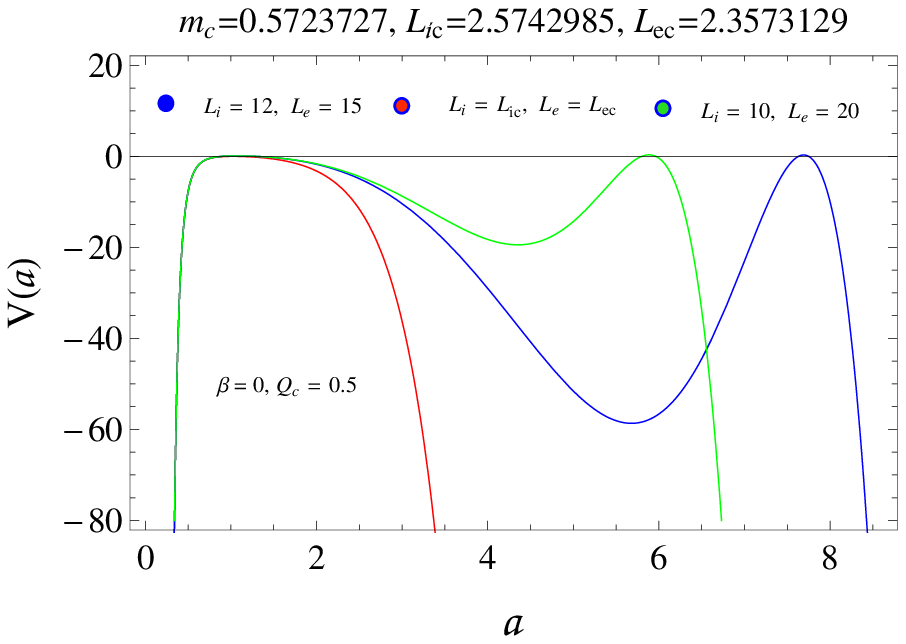,width=.5\linewidth}\epsfig{file=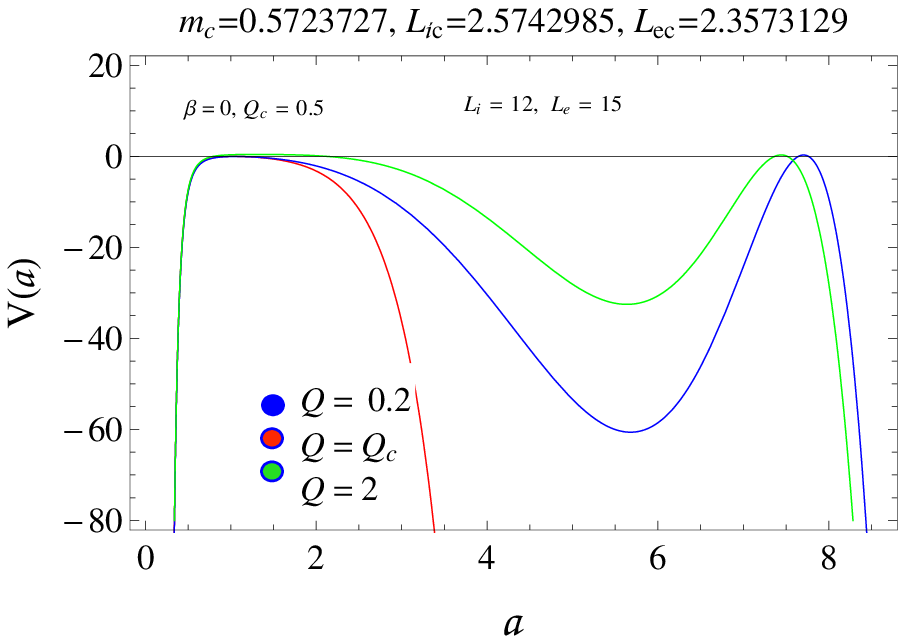,width=.5\linewidth}
\caption{The potential function for interior dS spacetime and
exterior Bardeen-dS BH with different values of $L_i$, $L_e$ (left
plot) and charge (right plot).}
\epsfig{file=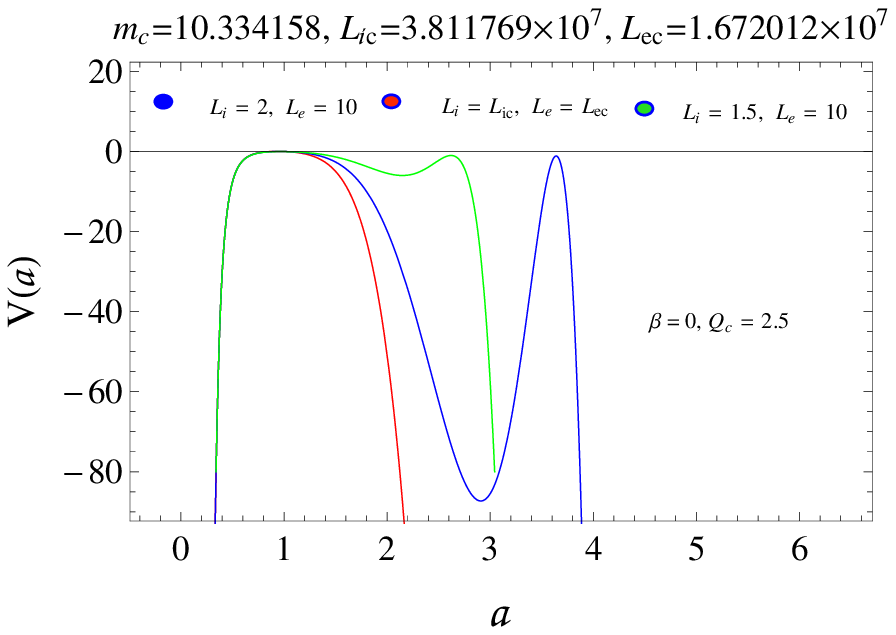,width=.5\linewidth}\epsfig{file=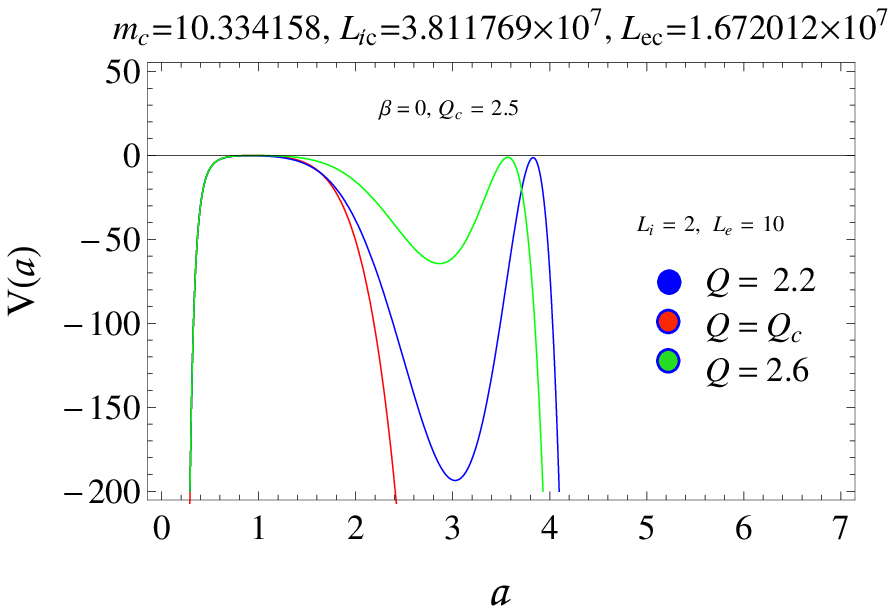,width=.5\linewidth}
\caption{The potential function for interior dS spacetime and
exterior Bardeen-dS BH with different values of $L_i$, $L_e$ (left
plot) and charge (right plot).}
\end{figure}

\subsection{Dust Fluid Shell}

Here, we explore the results in the presence of dust fluid
distribution at the shell. The corresponding expressions of the
potential function and its derivative can be written as
\begin{eqnarray}\nonumber
V(a)&=&\frac{1}{2}+\frac{a^6}{4 L_e^2 L_i^2}-\frac{a^6}{8
L_e^4}-\frac{a^6}{8 L_i^4}-\frac{a^2}{4 L_e^2}-\frac{a^2}{4
L_i^2}-\frac{a^2 m}{2 \left(a^2+Q^2\right)^{3/2}}-\frac{1}{8
a^2}\\\label{26}&-&\frac{a^6 m}{2 \left(a^2+Q^2\right)^{3/2}
L_e^2}+\frac{a^6 m}{2 \left(a^2+Q^2\right)^{3/2} L_i^2}-\frac{a^6
m^2}{2 \left(a^2+Q^2\right)^3},\\\nonumber V'(a)&=&\frac{3 a^5}{2
L_e^2 L_i^2}-\frac{3 a^5}{4 L_e^4}-\frac{3 a^5}{4 L_i^4}+\frac{1}{4
a^3}-\frac{a m}{\left(a^2+Q^2\right)^{3/2}}+\frac{3 a^7 m}{2
\left(a^2+Q^2\right)^{5/2} L_e^2}\\\nonumber&-&\frac{3 a^7 m}{2
\left(a^2+Q^2\right)^{5/2} L_i^2}+\frac{3 a^7
m^2}{\left(a^2+Q^2\right)^4}-\frac{3 a^5
m}{\left(a^2+Q^2\right)^{3/2} L_e^2}-\frac{3 a^5
m^2}{\left(a^2+Q^2\right)^3}\\\label{27}&+&\frac{3 a^5
m}{\left(a^2+Q^2\right)^{3/2} L_i^2}+\frac{3 a^3 m}{2
\left(a^2+Q^2\right)^{5/2}}-\frac{a}{2 L_e^2}-\frac{a}{2 L_i^2},
\end{eqnarray}
respectively. The respective critical values for different choices
of interior and exterior regions are expressed in Table
{\textbf{2}}. We choose suitable values of charge for which the
system shows real critical values of physical parameters.
\begin{table}[ht] \caption{Critical values of physical parameters and outcomes of
the developed structure for dust fluid shell with different internal
and external geometries.} \label{table:1} \vspace{0.5cm} \centering
\begin{tabular}{ |c c c c c c c c c |c| }
\hline \multicolumn{9}{|c|}{{Final outcomes of developed structure
for dust fluid shell ($\beta=1$)}}\\ \hline
$Q_c$ & $a_c$ & $m_c$ & $L_{ic}$ & $L_{ec}$ & IR & ER & St & Fig \\
\hline\hline
0.7 & 1.918218 & 0.891839 & - & - & FR & B & NS & \textbf{11}\\
0.8 & 2.981319 & 0.923782 & - & - & FR & B & NS & \textbf{12} \\
1.5 & 0.703307 & 3.778338 & - & - & FR & B & BH & \textbf{13} \\
\hline
0.7 & 1.021781 & 0.909516 & 6.64462$\times 10^7$ & - & dS & B & GS & \textbf{14}\\
0.8 & 2.981319 & 1.636369 & 15.37334 & - & dS & B & GS & \textbf{15} \\
1.5 & 12.79669 & 0.980800 & 4.15909$\times 10^9$ & - & dS & B & GS & \textbf{16} \\
\hline
0.7 & 1.918218 & 0.880891 & 8.692764 & 10.64767 & dS & BdS & GS & \textbf{17}\\
0.8 & 2.981319 & 0.916398 & 15.37334 & 17.40656 & dS & BdS & GS & \textbf{18} \\
1.5 & 12.79669 & 0.978446 & 109.3384 & 112.3030 & dS & BdS & GS & \textbf{19} \\
\hline
\end{tabular}
\end{table}

\subsubsection*{Case (i): $m\neq0\neq Q$ and
$\Lambda_i=0=\Lambda_e$}

The results of the collapse of a shell with an internal flat and
external Bardeen BH is shown in Figures \textbf{11-13}. The the
potential function represents the stable configuration of the
developed structure for every value of mass and charge which denotes
the presence of normal star (Figures \textbf{11} and \textbf{12}).
The shell shows collapsing behavior ($V(a)<0$) for small values of
shell radius and then shows expansion ($V(a)>0$). Figure \textbf{13}
indicates the collapsing configuration which leads to the formation
of BHs. There is no possible way to develop gravastar for every
choice of physical parameters.
\begin{figure}\centering\label{2}
\epsfig{file=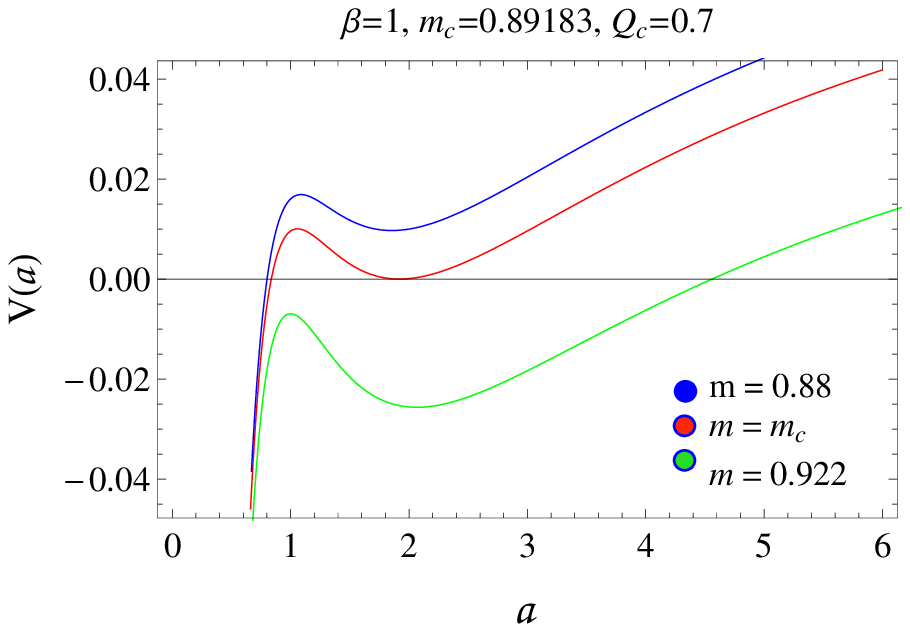,width=.5\linewidth}\epsfig{file=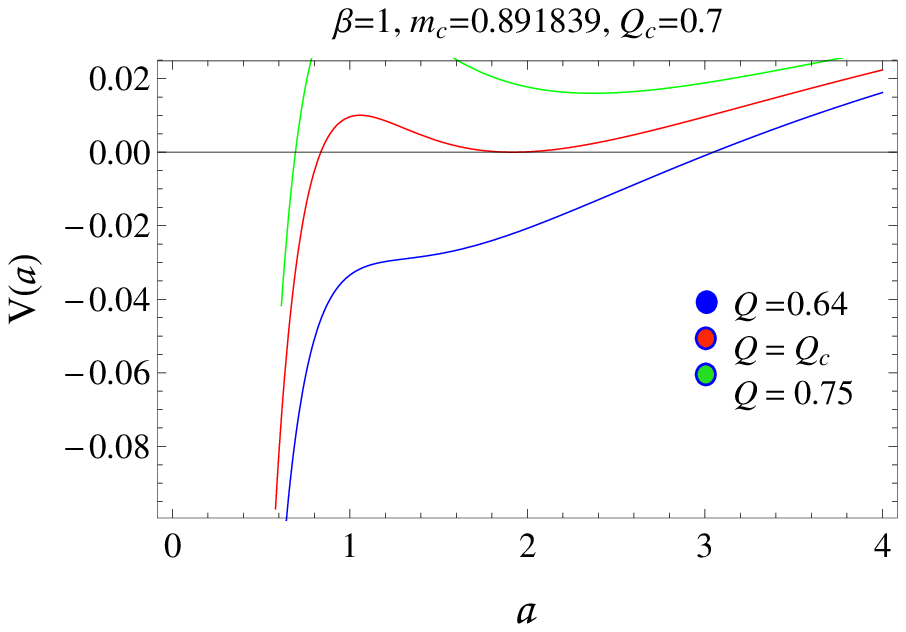,width=.5\linewidth}
\caption{The potential function for interior flat and exterior
Bardeen BH with different values of mass (left plot) and charge
(right plot).}
\epsfig{file=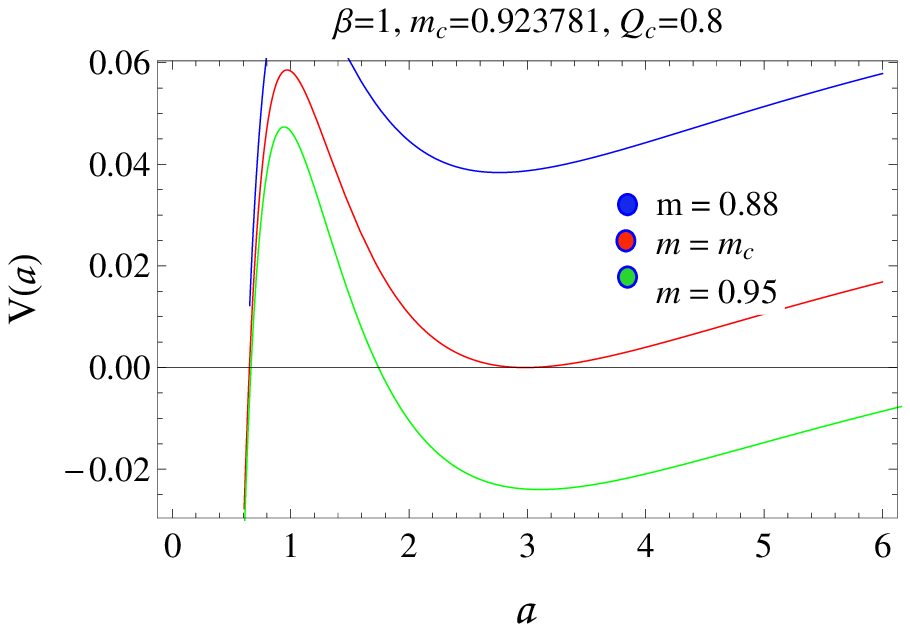,width=.5\linewidth}\epsfig{file=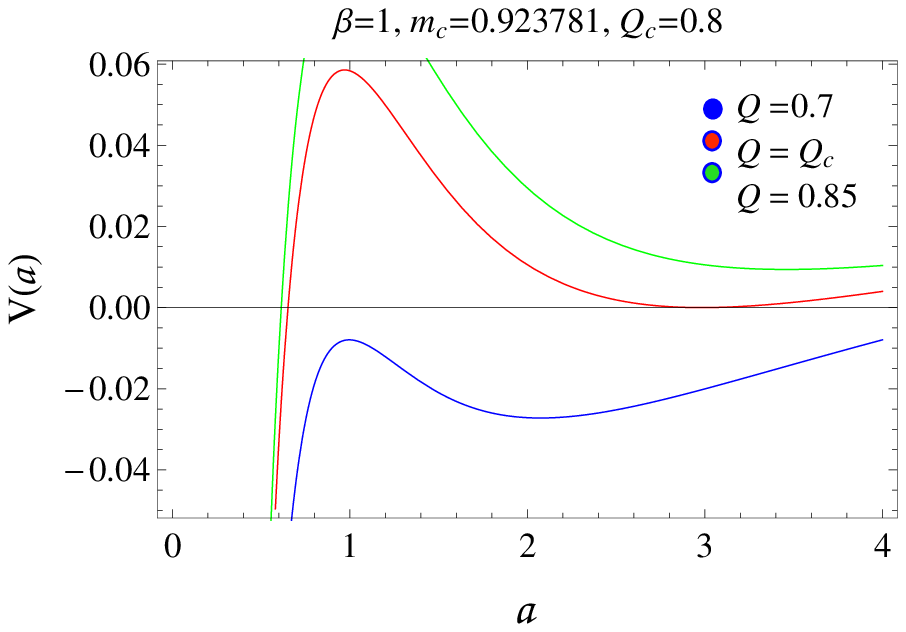,width=.5\linewidth}
\caption{The potential function for interior flat and exterior
Bardeen BH with different values of mass (left plot) and charge
(right plot).}
\epsfig{file=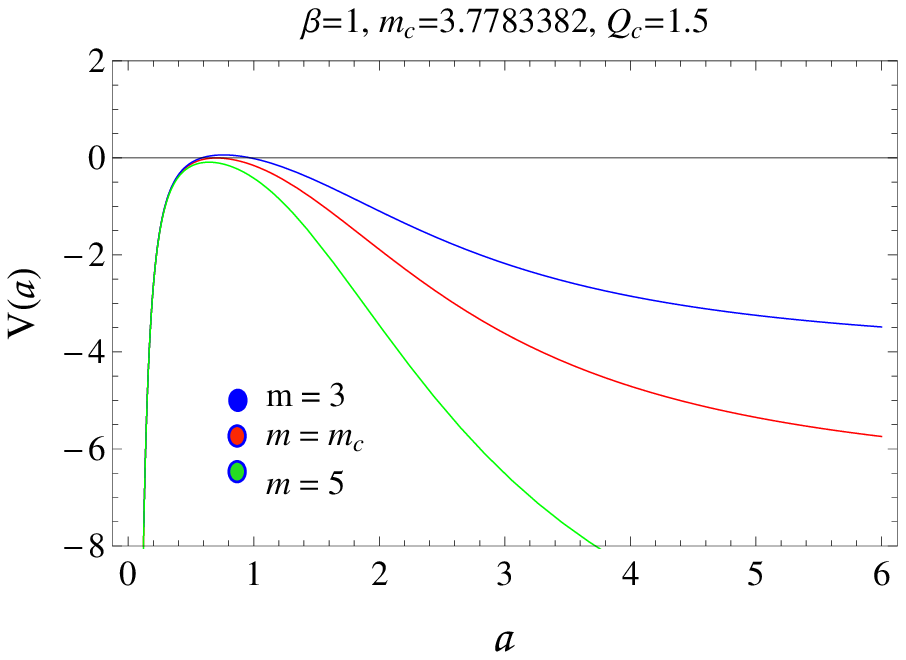,width=.5\linewidth}\epsfig{file=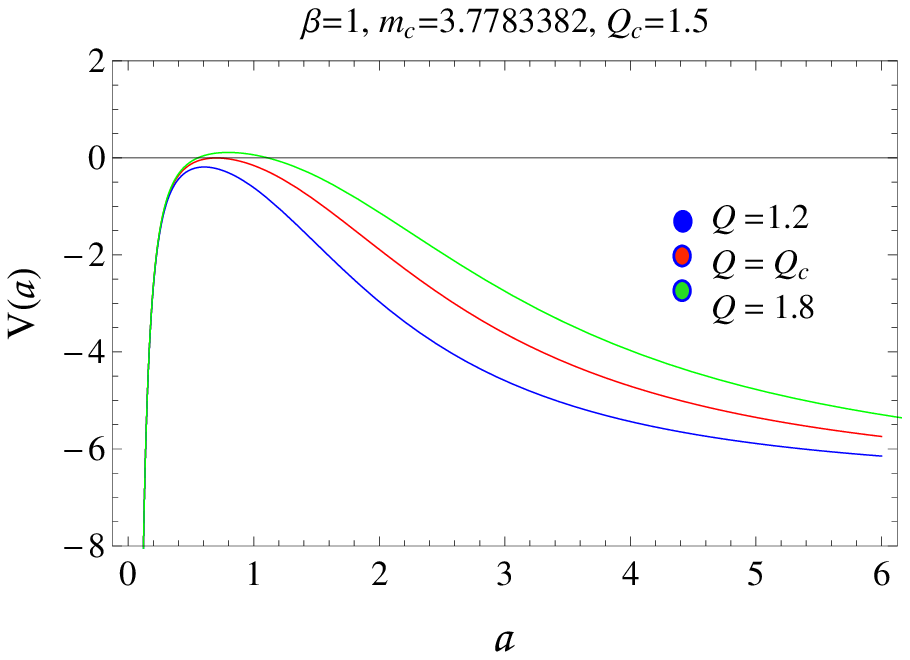,width=.5\linewidth}
\caption{The potential function for interior flat and exterior
Bardeen BH with different values of mass (left plot) and charge
(right plot).}
\end{figure}

\subsubsection*{Case (ii): $m\neq0\neq Q$, $\Lambda_i\neq0$ and
$\Lambda_e=0$}

Here we take the collapse of internal dS and external Bardeen BH
through critical values given in Table \textbf{2}. This is the most
suitable case for the existence of stable bounded excursion
gravastar. The stable gravastar must exist for the choice of
considered critical values (Table \textbf{2}) as shown in Figures
\textbf{14-16}. The potential function has two real roots as
$0<a_1<a_2$. If $a<a_1$, then the potential function approaches to
$-\infty$ as $a\rightarrow0$ that represents the flat spacetime. If
$a_1<a<a_2$, then the potential function is less than zero and hence
expresses the existence of stable bounded excursion gravastar. If
$a_2<a$, then $V(a)$ approaches to plus or minus infinity that
express the expansion and collapsing behavior as $a\rightarrow
\infty$. It is found that the choice of Bardeen BH as an exterior
geometry of shell filled with dust fluid provides stable bounded
excursion gravastar as compared to the choice of Schwarzschild/
Schwarzschild dS/RN geometries (Rocha et al 2008a; 2008b; Chan et al
2009b; 2010). Hence, this is more suitable for the construction of
stable gravastar with a dust shell.
\begin{figure}\centering\label{2}
\epsfig{file=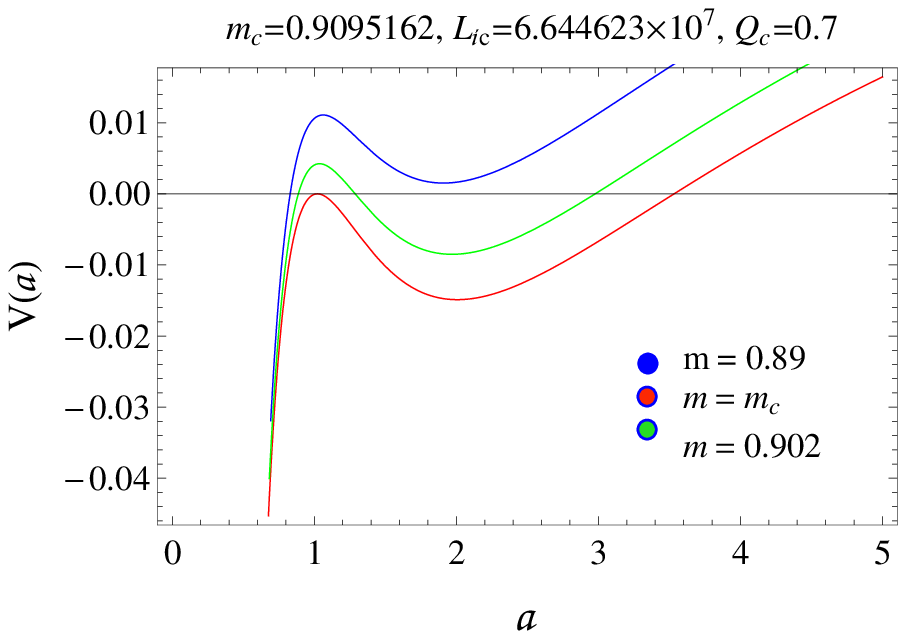,width=.5\linewidth}\epsfig{file=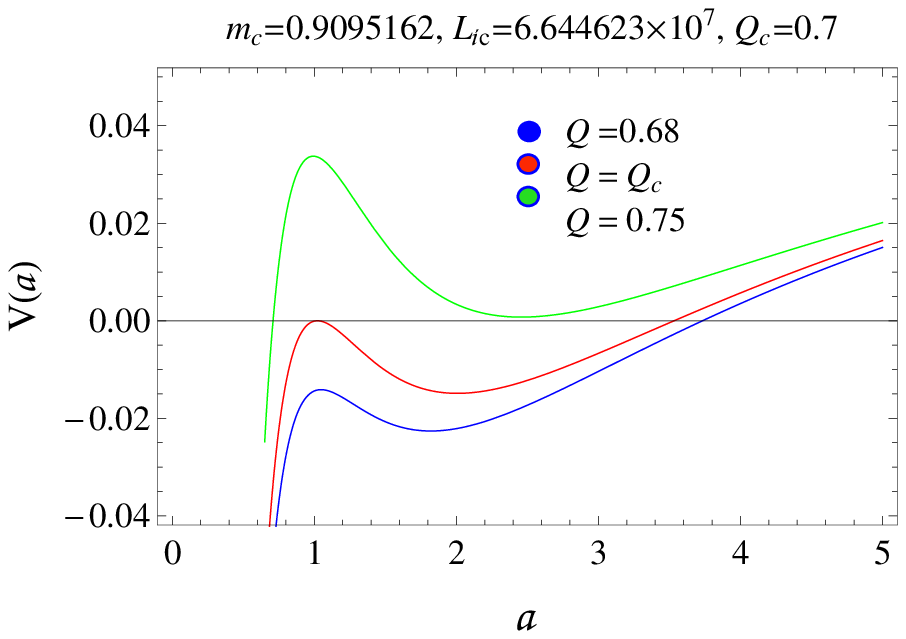,width=.5\linewidth}
\caption{The potential function for interior dS spacetime and
exterior Bardeen BH with different values of mass (left plot) and
charge (right plot).}
\epsfig{file=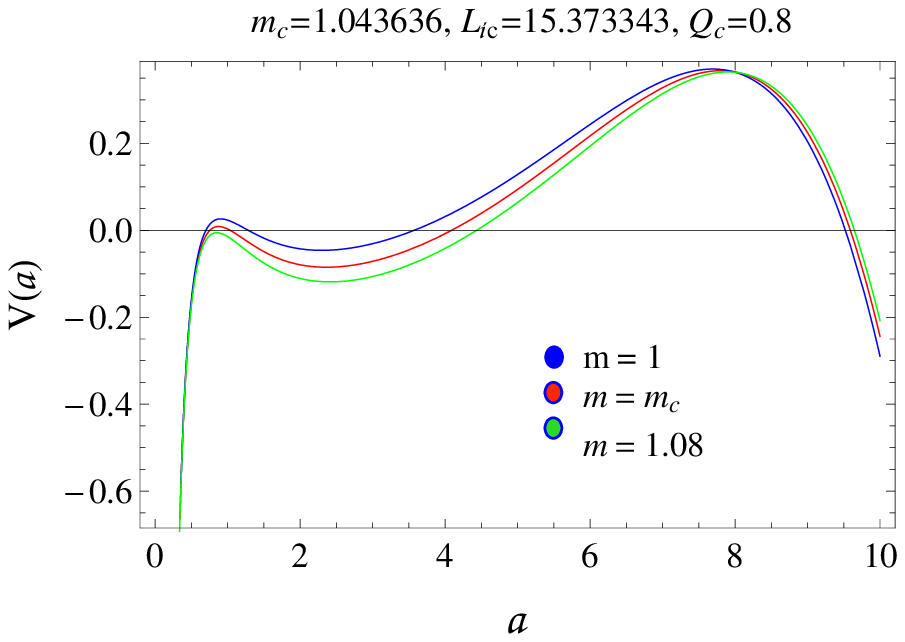,width=.5\linewidth}\epsfig{file=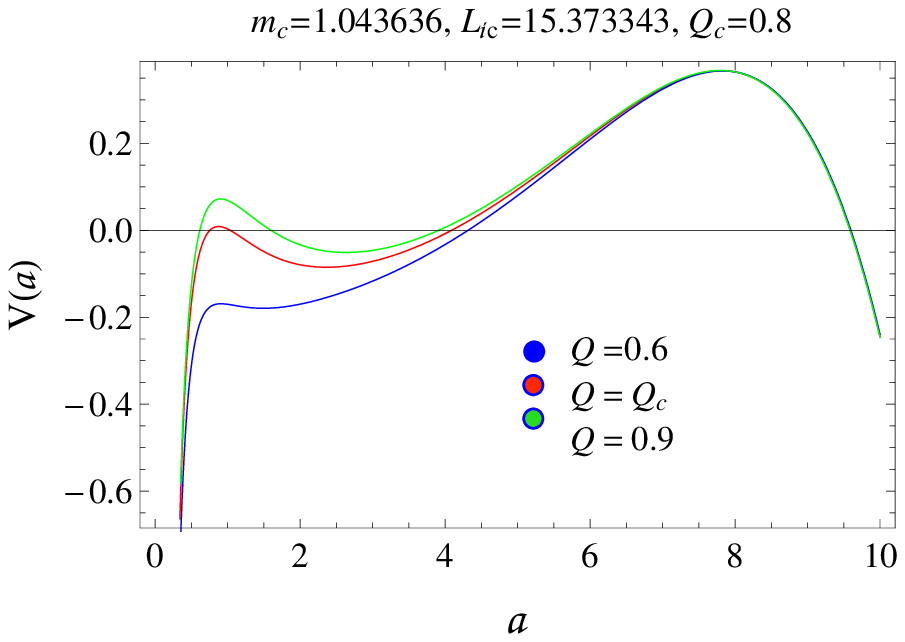,width=.5\linewidth}
\caption{The potential function for interior dS spacetime and
exterior Bardeen BH with different values of mass (left plot) and
charge (right plot).}
\epsfig{file=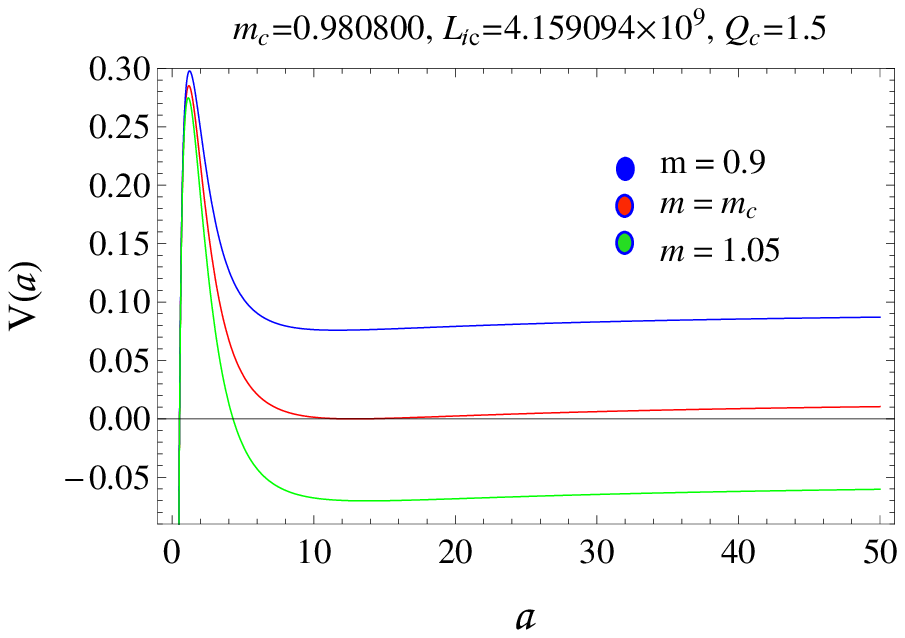,width=.5\linewidth}\epsfig{file=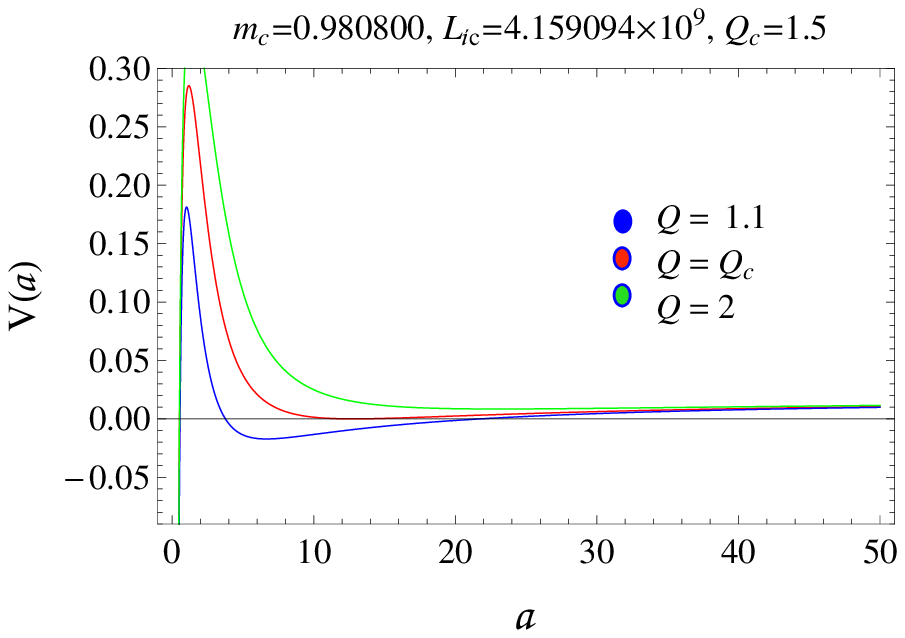,width=.5\linewidth}
\caption{The potential function for interior dS spacetime and
exterior Bardeen BH with different values of mass (left plot) and
charge (right plot).}
\end{figure}

\subsubsection*{Case (iii): $m\neq0\neq Q$ and
$\Lambda_i\neq0\neq\Lambda_e$}
\begin{figure}\centering\label{2}
\epsfig{file=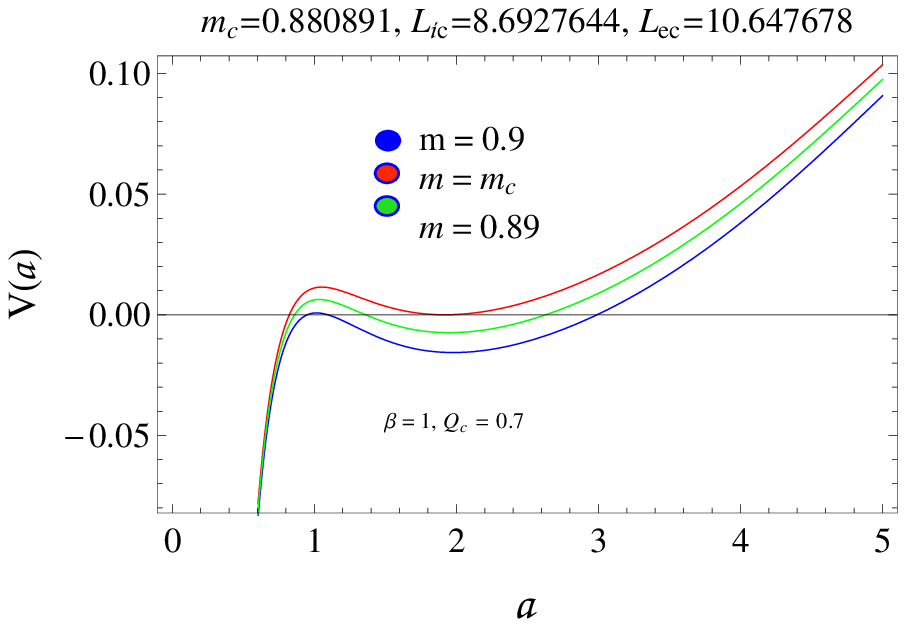,width=.5\linewidth}\epsfig{file=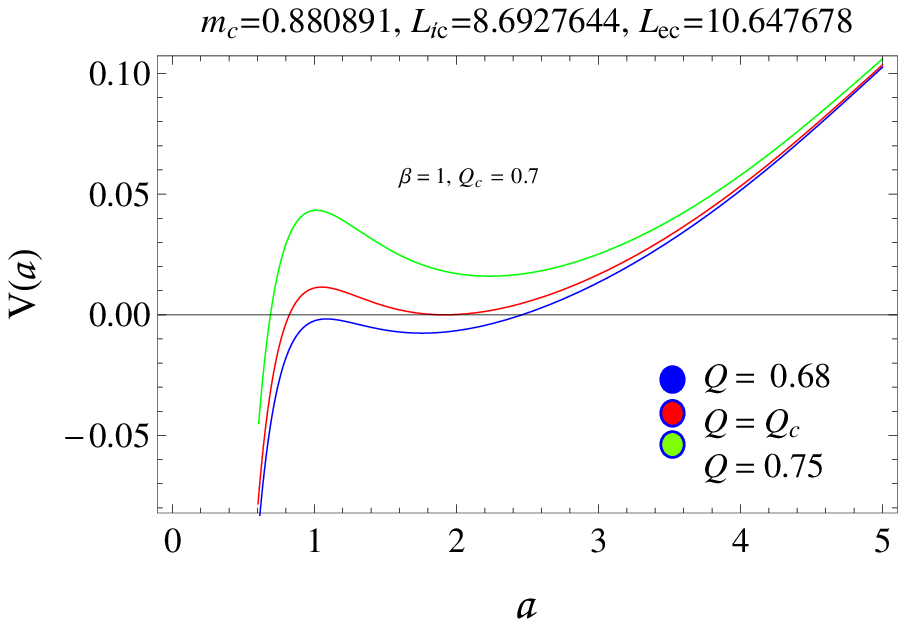,width=.5\linewidth}
\caption{The potential function for interior dS spacetime and
exterior Bardeen-dS BH with different values of mass (left plot) and
charge (right plot).}
\epsfig{file=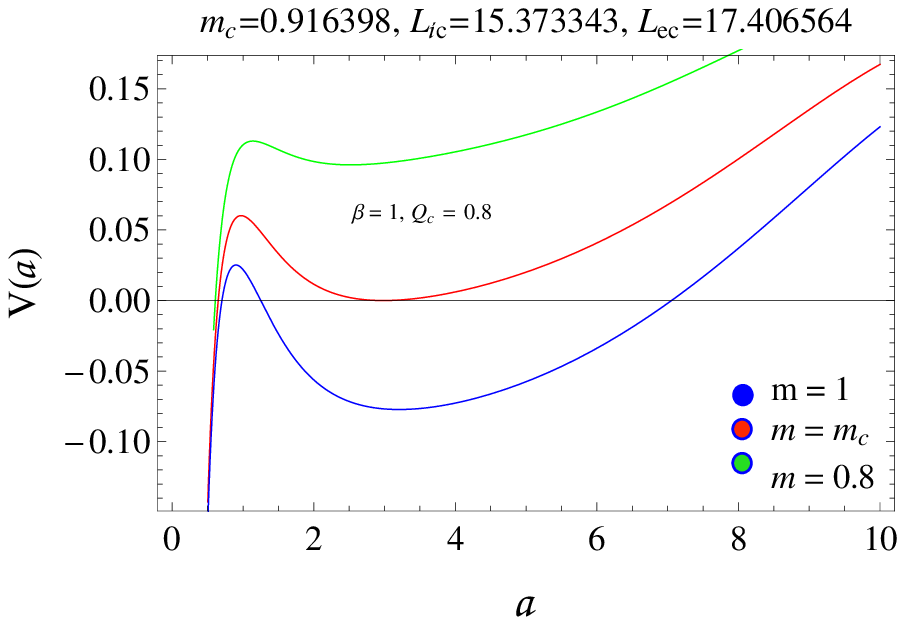,width=.5\linewidth}\epsfig{file=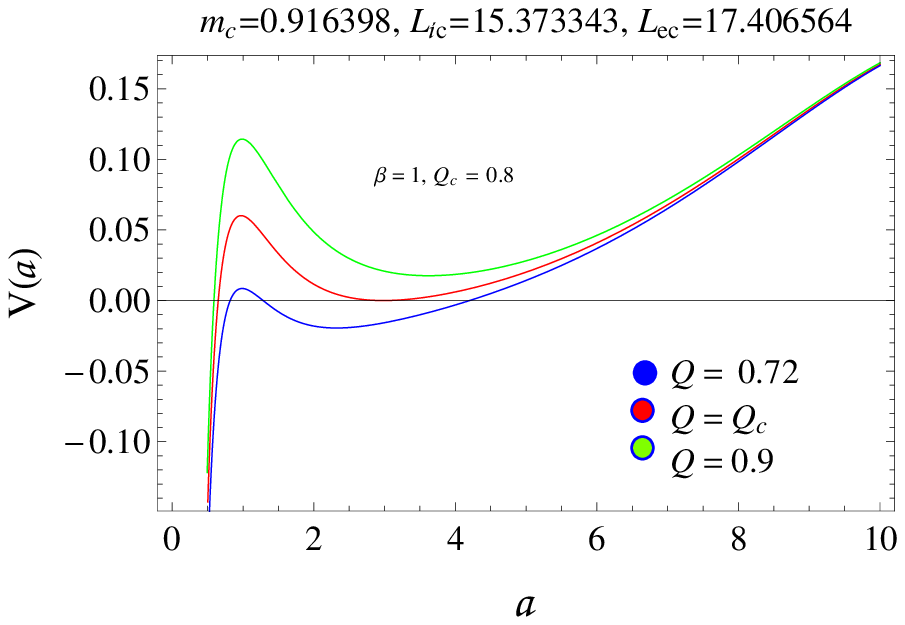,width=.5\linewidth}
\caption{The potential function for interior dS spacetime and
exterior Bardeen-dS BH with different values of mass (left plot) and
charge (right plot).}
\epsfig{file=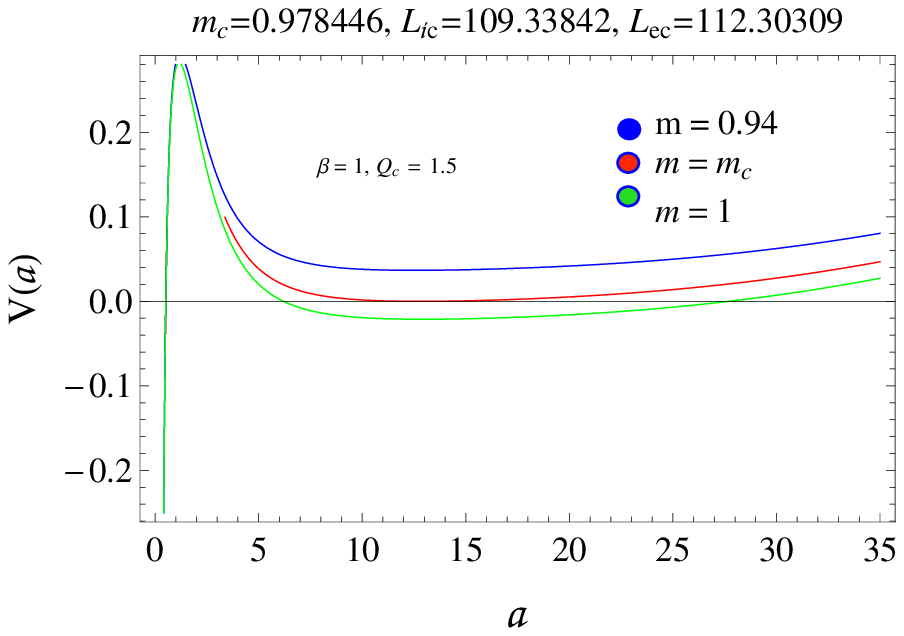,width=.5\linewidth}\epsfig{file=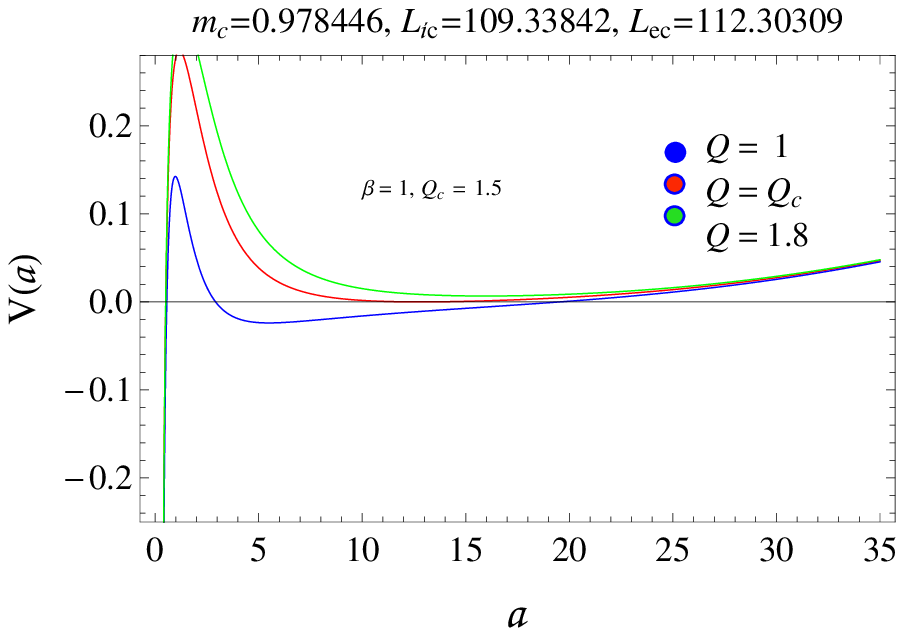,width=.5\linewidth}
\caption{The potential function for interior dS spacetime and
exterior Bardeen-dS BH with different values of mass (left plot) and
charge (right plot).}
\end{figure}

We consider the internal dS region and external Bardeen-dS BH. The
critical values (\textbf{Table 2}) follow the inequality $L_e>L_i$
with every considered choice of $Q$. Hence, the developed structure
is suitable to eliminate the presence of an event horizon and
singularity for critical values. The graphical behavior of the
potential function shows the expanding behavior which gives
information about the existence of stable gravastar near the
suitable critical values of physical parameters (Figures
\textbf{17-19}). Hence, the stable gravastar must exist for the
different choices of $m$ and $Q$.

\section{Final Remarks}

This work is the extension of stability issues related to the
possible existence of stable bounded excursion gravastar from a
model consisting of internal dS region, intermediate thin-shell
filled with matter distribution followed the EoS $p=(1-\beta)\rho$
and external Schwarzschild dS as well as RN spacetimes (Rocha et al
2008a; 2008b; Chan et al 2009b; 2010). Here, we are interested to
explore the existence of bounded excursion gravastar in the
background of exterior regular Bardeen and Bardeen-dS BHs. We have
used the cut and paste approach to match these spacetimes at the
shell. The potential function of the shell filled with matter
distribution ($p=(1-\beta)\rho$) can be evaluated through the
equation of motion of the shell. The developed framework is
investigated through the critical values of physical parameters by
solving the following equations $V(a)=0$ and $V'(a)=0$,
simultaneously. For this purpose, we have considered two types of
matter distribution, stiff ($\beta=0$) and dust ($\beta=1$) fluids.

For the stiff fluid shell, the critical values of physical
parameters for three different cases are given in Table \textbf{1}.
For the interior flat spacetime and exterior Bardeen BHs, we have
obtained the collapsing behavior of the shell which indicates the
formation of BHs (Figures \textbf{2-4}). The interior dS and
exterior Bardeen BH represent the region of stable bounded excursion
gravastar for a suitable choice of physical parameters. If the
gravastar model exists then there is a possibility for the existence
of BH and vice versa. We have analyzed that the regions of stable
bounded excursion gravastar increase by increasing mass and $L_i$
(Figures \textbf{5-7}). For the interior dS and exterior Bardeen-dS
BH, we have obtained that bounded excursion stable gravastar must
exist if $L_i<L_e$ (Figures \textbf{8-10}). The regions of bounded
excursion gravastar increase by enhancing charge and $L_e$.

For dust fluid shell, we have obtained the respective critical
values of physical parameters (Table \textbf{2}). There does not
exist the possibility of the existence of gravastar for the flat and
exterior Bardeen BH (Figures \textbf{11-13}). The stable structure
of bounded excursion gravastar is found for every choice of physical
parameters for the interior dS and exterior Bardeen BH (Figures
\textbf{14-16}). The stable bounded excursion gravastar is
determined for suitable choices of physical parameters with interior
dS and exterior Bardeen-dS BH (Figures \textbf{17-19}).

We conclude that the gravastar model in the background of regular
BHs is more feasible for the existence of a stable configuration of
bounded excursion gravastar than the Schwarzschild and RN BHs (Rocha
et al 2008a; 2008b; Chan et al 2009b; 2010). These BHs do not
provide any region of stable gravastar for the choice of dust shell
while our models show stable regions of bounded excursion gravastar.
This paper also follows the previous results for the choice of $Q=0$
(Chan et al 2009b).\\\\
\textbf{Conflict of Interest Statement:} The authors declare that
they do not have any conflict of interest.

\end{document}